\documentclass[reprint,superscriptaddress,amssymb,amsmath,aps,prx,longbibliography]{revtex4-1}

\usepackage{graphicx}
\usepackage{amssymb}
\usepackage{epstopdf}
\usepackage{upgreek}
\usepackage{dcolumn}
\usepackage{bm}
\usepackage{gensymb}
\usepackage{braket}
\usepackage{amsmath}
\usepackage{mathtools}
\usepackage{float}

\usepackage{tikz}

\expandafter\ifx\csname package@font\endcsname\relax\else
\expandafter\expandafter
\expandafter\usepackage
\expandafter\expandafter
\expandafter{\csname package@font\endcsname}%
\fi

\newcommand{\micro}{$\upmu$}

\newcommand{\be}{\begin{eqnarray}}
\newcommand{\ee}{\end{eqnarray}}
\newcommand{\bfig}{\begin{figure}}
	\newcommand{\efig}{\end{figure}}

\DeclareFontFamily{U}{mathb}{}
\DeclareFontShape{U}{mathb}{m}{n}{
	<-5.5> mathb5
	<5.5-6.5> mathb6
	<6.5-7.5> mathb7
	<7.5-8.5> mathb8
	<8.5-9.5> mathb9
	<9.5-11.5> mathb10
	<11.5-> mathbb12
}{}

\usepackage[colorlinks=true,allcolors=blue]{hyperref}%

\begin{document}
	
	\title{Squeezed vacuum used to accelerate the search for a weak classical signal}
	\author{M. Malnou}
	\thanks{These two authors contributed equally}
	\email{maxime.malnou@colorado.edu}
	\affiliation{JILA, National Institute of Standards and Technology and the University of Colorado, Boulder, Colorado 80309, USA}
	\affiliation{Department of Physics, University of Colorado, Boulder, Colorado 80309, USA}
	\author{D. A. Palken}
	\thanks{These two authors contributed equally}
	\email{maxime.malnou@colorado.edu}
	\affiliation{JILA, National Institute of Standards and Technology and the University of Colorado, Boulder, Colorado 80309, USA}
	\affiliation{Department of Physics, University of Colorado, Boulder, Colorado 80309, USA}
	\author{B. M. Brubaker}
	\affiliation{JILA, National Institute of Standards and Technology and the University of Colorado, Boulder, Colorado 80309, USA}
	\affiliation{Department of Physics, University of Colorado, Boulder, Colorado 80309, USA}
	\author{Leila R. Vale}
	\affiliation{National Institute of Standards and Technology, Boulder, Colorado 80305, USA}
	\author{Gene C. Hilton}
	\affiliation{National Institute of Standards and Technology, Boulder, Colorado 80305, USA}
	\author{K. W. Lehnert}
	\affiliation{JILA, National Institute of Standards and Technology and the University of Colorado, Boulder, Colorado 80309, USA}
	\affiliation{Department of Physics, University of Colorado, Boulder, Colorado 80309, USA}
	\date{\today}
	
	\date{\today}
	
	\begin{abstract}
		 Many experiments that interrogate fundamental theories require detectors whose sensitivities are limited by the laws of quantum mechanics. In cavity-based searches for axionic dark matter, vacuum fluctuations in the two quadratures of the cavity electromagnetic field limit the sensitivity to an axion-induced field. In an apparatus designed to partially mimic existing axion detectors, we demonstrate experimentally that such quantum limits can be overcome through the use of squeezed states. By preparing a microwave cavity in a squeezed state and measuring just the squeezed quadrature, we enhance the spectral scan rate by a factor of $2.12 \pm 0.08$. This enhancement is in excellent quantitative agreement with a theoretical model accounting for both imperfect squeezing and measurement. 
	\end{abstract}
	
	\maketitle

	\section{INTRODUCTION}    
    In searches for dark matter axions \cite{peccei1977cp,peccei1977constraints,preskill1983cosmology,abbott1983a,dine1983the}, the quadratures $\hat X$ and $\hat Y$ of a resonant cavity's electromagnetic field carry the imprint of the dark matter signal as a slight excess in their power spectra. Problematically, the quantum noise \cite{caves1982quantum} intrinsic to a measurement of these observables overpowers the signal by roughly three orders of magnitude, and the signal frequency is \textit{a priori} unknown \cite{bradley2003microwave}. Even at the quantum limit,  scanning the 1--10 GHz frequency band at the pessimistic benchmark DFSZ \cite{zhitnitsky1980possible, dine1981simple} coupling with present detector technologies \cite{du2018search,zhong2018results} will take upwards of 20,000 years of experimental live time. Realizing the several thousandfold speed-up required to put these detectors on expedient schedules will require many parallel innovations in detector design and sensitivity. Among the most promising advances are those which allow access to a fundamentally distinct parameter regime in which axion searches are no longer limited by quantum noise. 
	
    In detectors measuring the quadratures of a resonant mode, quantum noise can be circumvented by preparing the mode in a squeezed state \cite{caves1981quantum}. Squeezing unbalances the uncertainties in the two quadrature observables, thereby permitting precise knowledge of one at the expense of the other. Leveraging quantum squeezing to enhance measurement sensitivity has been a long-standing goal in the optical domain for the sensing of gravitational waves \cite{caves1981quantum, abadie2011gravitational, aasi2013enhanced, kimble2001conversion}. At microwave frequencies, meanwhile, squeezing has been demonstrated in principle \cite{murch2013reduction, clark2016observation, bienfait2017magnetic}, but it has yet to aid in a search for new physical phenomena.
    
    But is microwave squeezing beneficial in an axion dark matter search given current experimental constraints? This question involves both matters of principle and practice. Specifically, in axion haloscope \cite{sikivie1983experimental} searches, the dark matter signal is a persistent tone of unknown frequency and phase, but more coherent than the resonant cavity itself. As such, a haloscope uses a tunable cavity to scan a resonance through frequency, with the critical parameter being the rate at which the cavity can be tuned with some specified sensitivity to an axion. The suitability of squeezing to this type of search has not received specific theoretical attention. Furthermore, if squeezing is in principle beneficial in this case, can microwave losses be made sufficiently low to yield a practical improvement?

    In this article, we show both theoretically and experimentally that squeezing increases the scan rate in a search for a weak, axion-like signal of unknown frequency. This improvement exists in spite of the fact that squeezing does not improve the sensitivity of a haloscope to a tone of known frequency. We demonstrate this speed-up in an apparatus designed to mimic the behavior of existing haloscopes. By delivering a maximum of $4.5\pm 0.1$ dB of squeezing from one Josephson parametric amplifier (JPA)  \cite{yamamoto2008flux,castellanos2008amplification,zhou2014high} to another in a squeezed state receiver (SSR) configuration, we perform a realistic acquisition and processing protocol both with and without squeezing enabled. We demonstrate that the SSR accelerates the scan rate at constant sensitivity to axion-like signals by a factor of $2.12\pm0.08$. Equipping the SSR to a dark matter detector such as HAYSTAC \cite{brubaker2017first, zhong2018results} or ADMX \cite{du2018search} will mark the transition of searches for physics beyond the standard model to the sub-quantum limited noise regime.


	\section{THEORY OF SQUEEZING-ENHANCED SCAN RATE}
	\label{quantsq}
\begin{figure}[!htpb]
		\centering
		\includegraphics[scale=0.72]{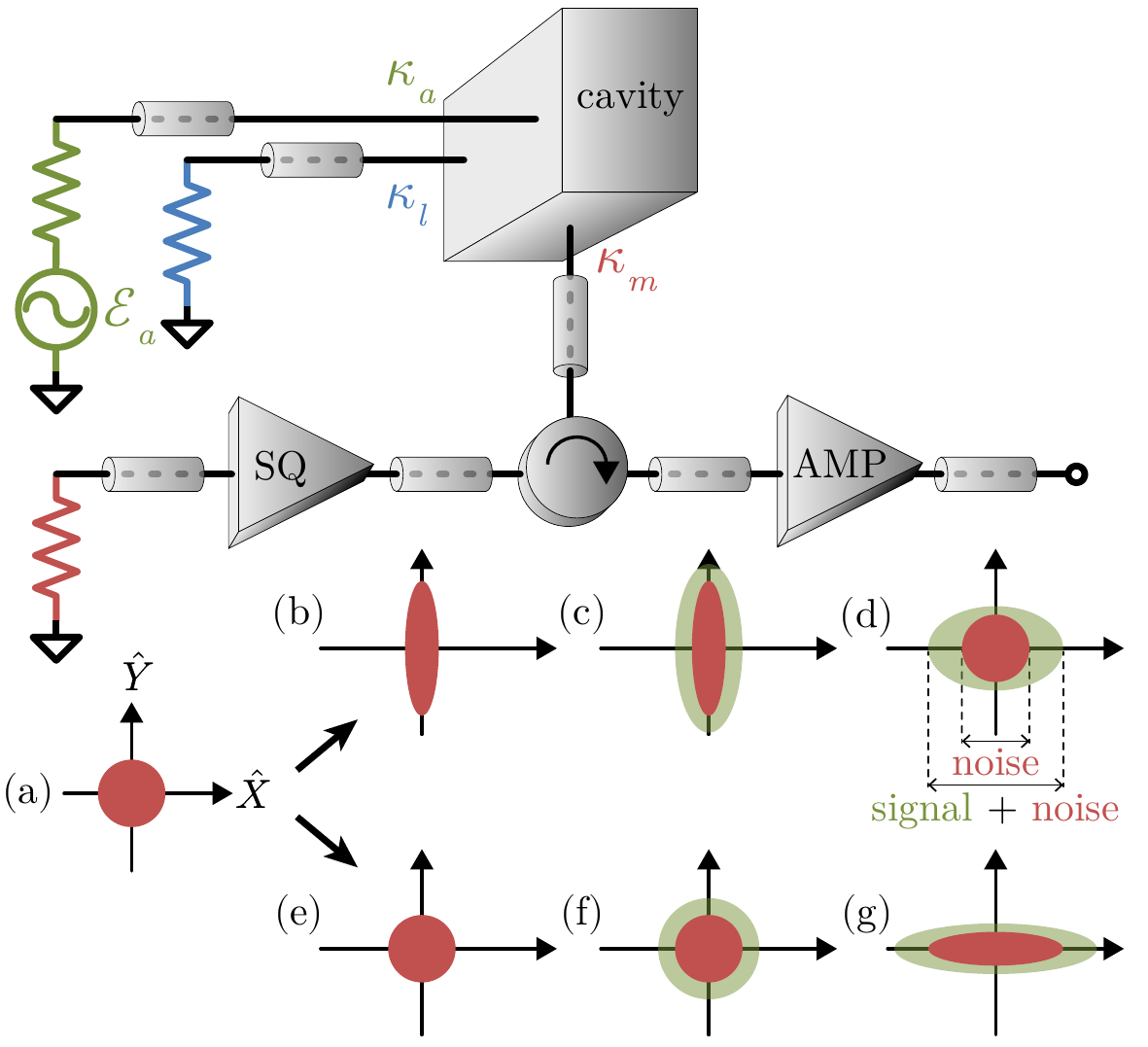}
		\caption{Schematic of the SSR and cavity. Two JPAs, SQ and AMP, respectively squeeze and read out a microwave field interacting with a cavity at rate $\kappa_m$. An axion-like field $\mathcal{E}_a$, coupled to the cavity at rate $\kappa_a$, displaces the cavity state. Energy leaves the cavity through internal absorption at a rate~$\kappa_l$. Bottom panel: quadrature representation of a vacuum state detuned from the cavity resonant frequency, traveling in the SSR. At the SQ's input (a) it is Gaussian, azimuthally equiprobable in the $(\hat{X},\hat{Y})$ plane (red disk). The state is squeezed along $\hat{X}$ by the SQ (b), displaced along a random phase within the cavity (c), and amplified along $\hat{X}$ by the AMP (d). Comparing to what happens without squeezing (e)--(g), the size of the signal-plus-noise (green) relative to the noise (red) in this quadrature is larger with (d) than without (g) squeezing.}
		\label{fig:schematic}
	\end{figure}

	Figure~\ref{fig:schematic} shows a representative experimental apparatus in which a resonant cavity is coupled to an SSR comprising a pair of JPAs. We start by considering the cavity, whose internal mode (frequency $\omega_c$) has Hamiltonian $\hat{H} = \hbar\omega_c(\hat{X}^2 + \hat{Y}^2)/2$, where $\hat{X}$ and $\hat{Y}$ are the noncommuting quadratures of the cavity field, obeying $[\hat X, \hat Y] = i$. This cavity is modeled as exchanging energy with three ports. First, a measurement port couples the cavity mode to the propagating modes of a transmission line with power decay rate $\kappa_m$. Along this line, a microwave circulator spatially separates incoming and outgoing propagating modes. Second, a loss port, connected to a fictitious transmission line, models the cavity's internal energy dissipation at rate $\kappa_l$. Third, the cavity's coupling to the signal of interest at rate $\kappa_a$ is modeled as occurring through another fictitious transmission line; the signal itself is modeled as a  microwave generator characterized by its frequency $\omega_a$, spectral width $\Delta_a$, and amplitude $\mathcal{E}_a$. We assume $\mathcal{E}_a \gg 1$, implying that the displacement of the cavity mode by the signal is classical (i.e., the contribution of the signal to the cavity's quantum fluctuations can be neglected in comparison to those coming from the loss and measurement ports). We also assume a narrow band signal ($\Delta_a \ll \{\kappa_l,\kappa_m\}$) so weakly coupled ($\mathcal{E}_a^2 \kappa_a \ll \kappa_l$) that the time required to resolve the displacement is much longer than the signal's phase coherence time. Therefore, on average, this displacement yields a small excess power above the vacuum fluctuations, isotropic in quadrature space (see bottom panel in Fig.\,\ref{fig:schematic}). These inequalities are very well satisfied in the case of the axion field (see Appendix~\ref{sup:axionmodel}).
	
    The SSR itself comprises the two JPAs shown in Fig.\,\ref{fig:schematic}, which couple respectively to incoming and outgoing modes at the cavity's measurement port. This configuration exploits the fact that a portion of the vacuum noise exiting the measurement port arises from vacuum noise incident on that same port. The first JPA (called SQ) squeezes these input fluctuations along the $\hat{X}$ quadrature, reducing the observable's variance below vacuum levels: $\sigma_X^2 < 1/2$ \cite{mallet2011quantum,bienfait2017magnetic,menzel2012path, boutin2017effect,malnou2018optimal}. To satisfy the uncertainty principle, the opposing quadrature's fluctuations exceed vacuum, $\sigma_Y^2 > 1/2$, but are not measured. The squeezed input field subsequently enters the cavity, where a small displacement by the signal would yield a small excess power in both quadratures. At the measurement port output, the second JPA (called AMP) noiselessly amplifies only the $\hat{X}$ quadrature with sufficient gain to overwhelm the noise added by following amplifiers and mixers \cite{yamamoto2008flux, zhou2014high, pogorzalek2017hysteretic}. Note that in the absence of squeezing there is neither a benefit nor a penalty associated with measuring only one quadrature compared to the usual two-quadrature case (see Appendix~\ref{sup:singledouble}). 
    
  
    The benefit of squeezing can be understood by analyzing the microwave network formed by the combination of SSR and cavity using input-output theory (see Appendix~\ref{sup:SSR}). The signal spectral density at the measurement port output is equal to the signal spectral density at the measurement port input, weighted by the susceptibility of the measurement port output to the signal port input. Similarly, the noise density at the measurement port output is a susceptibility-weighted sum of squeezed and unsqueezed noise from the measurement and loss ports, respectively. In the absence of transmission losses between the JPAs and the cavity, the ratio of output signal spectral density to total output noise spectral density (hereafter called the signal visibility) is thus
	\begin{equation} \label{SNR}
    \alpha(\omega) \approx
    \frac{n_A\kappa_a\kappa_m}{\left(n_T + \frac{1}{2}\right)\left(\kappa_l\kappa_m + \frac{\beta(\omega)}{G_s}\right)}.
	\end{equation}
    Here, $\omega$ is the frequency relative to cavity resonance, $n_T = 1/[\exp{(\hbar\omega_c/k_BT)} - 1]$ is the mean thermal photon number incident on the cavity from the measurement and loss ports, $n_A = \mathcal{E}_a^2$ is the mean photon number sourced by the fictitious generator, $\beta(\omega) = (\kappa_m - \kappa_l)^2/4 + \omega^2$, and $G_s$ is the power gain of the SQ (ideally equal to the reduction of the squeezed state variance below the vacuum value).
	
	When optimizing $\alpha(\omega)$ in Eq.\,\eqref{SNR}, we assume that $\kappa_m$ and $G_s$ can be freely varied, whereas $n_T$, $n_A$, $\kappa_l$, and $\kappa_a$ are fixed by the physics of the signal source or technical constraints of the detector. On cavity resonance ($\omega=0$), $\alpha$ is maximized at critical coupling ($\kappa_m = \kappa_l$). Because $\beta(0) = 0$ at critical coupling, $\alpha(0)$ is independent of $G_s$ and there is no benefit from squeezing; physically, the squeezed state injected into the cavity is completely absorbed in it, while all the unsqueezed noise from the loss port reaches the AMP. For $\omega \neq 0$, squeezing increases $\alpha(\omega)$ for any value of $\kappa_m$, as $G_s$ reduces the amount of measurement port noise reaching the output. In the limit where $G_s\rightarrow\infty$, the $\beta(\omega)$ term in the denominator can be neglected, and $\alpha(\omega)$ approaches the critically coupled resonant value $\alpha(\omega=0,\kappa_m=\kappa_l)$ for \textit{any} $\kappa_m$ and \textit{all} $\omega$. This illustrates an important point of principle: squeezing cannot improve the peak sensitivity of a haloscope, but there is no \textit{fundamental} limit to how much it can enhance the detector bandwidth over which this peak sensitivity is achieved. When $G_s$ is finite, overcoupling ($\kappa_m > \kappa_l$) increases the cavity bandwidth at the cost of reducing $\alpha(0)$. This can be a favorable trade-off (even in the absence of squeezing) because the signal's frequency $\omega_a$ is \textit{a priori} unknown \cite{irastorza2018new, chaudhuri2018fundamental}, and broader bandwidth enables larger cavity tuning steps. Moreover, squeezing mitigates the reduction of $\alpha(0)$ from overcoupling, thus enabling faster tuning without significant degradation of sensitivity.
   	
	To quantify this speed-up, we calculate the rate $R$ (in Hz/s) at which we can tune the cavity resonance through frequency space in search for a signal.
	This scan rate is inversely proportional to the measurement time at each tuning step, which in turn scales with $\alpha^{-2}$ as a consequence of Gaussian noise statistics (see Appendix~\ref{sup:SSRnoloss}). Hence $R\propto\Delta_a\int_{-\infty}^\infty\alpha^2(\omega) d\omega$; carrying out the integral we obtain
	\begin{equation} \label{R}
	R \propto \frac{\Delta_a\sqrt{G_s}n_A^2\kappa_a^2\kappa_m^2}{\left(n_T+\frac{1}{2}\right)^2\left[\kappa_l\kappa_m + \frac{1}{G_s}\left(\frac{\kappa_l-\kappa_m}{2}\right)^2\right]^{3/2}}.
	\end{equation}
	Without squeezing ($G_s=1$), $R$ is maximized when the cavity is twice overcoupled ($\kappa_m=2\kappa_l$), and at this optimal coupling the scan rate scales as $R_u^\mathrm{max}\propto\kappa_a^2/\kappa_l$, a known result \cite{alkenany2017design}. When $G_s\gg1$, the optimal coupling is $\kappa_m=2G_s\kappa_l$ and the scan rate scales as  $R_s^\mathrm{max}\propto G_s\kappa_a^2/\kappa_l$. Comparing the two situations, the scan rate is improved by $G_s$, which shows that an ideal SSR greatly accelerates the search for a weak classical signal when squeezing and overcoupling.	
	
	In practice, however, losses in microwave components reduce the SQ-to-AMP transmission efficiency $\eta$ and hence the benefit of squeezing, because part of the squeezed state is replaced with unsqueezed vacuum (see Appendix~\ref{sup:SSRwloss}). Figure~\ref{fig:Rtheo} compares the theoretical scan rate enhancement $E_t=R_s/R_u^\mathrm{max}$ (a) when $\eta=1$ (perfect transmission), and (b) when $\eta=0.69$ (efficiency we observe in practice), as a function of $G_s$ and $\kappa_m/\kappa_l$. In the first case, $E_t$ improves arbitrarily as squeezing and coupling are together increased. In the second case, it plateaus at $E_t^\mathrm{max}\approx2.2$ for $G_s>20$ when optimally coupled. In Secs.\, \ref{scanrate} and \ref{sec:faxion detection}, we present experimental results for the scan rate enhancement consistent with this theory.
	\begin{figure}[!h] 
		\centering
		\includegraphics[scale=0.185]{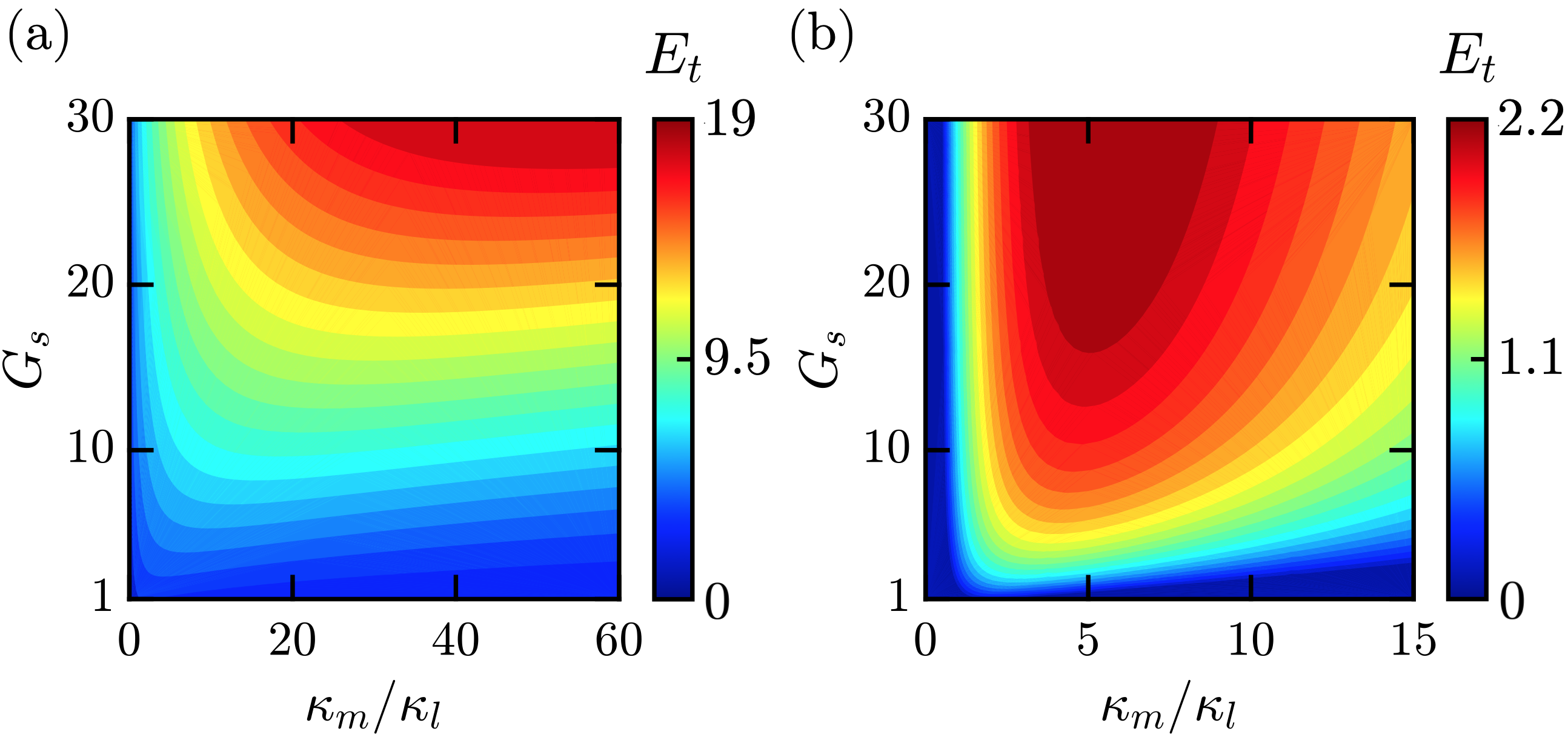}
		\caption{The scan rate enhancement $E_t$, calculated as a function of SQ single-quadrature power gain $G_s$ and coupling ratio $\kappa_m/\kappa_l$. With perfect efficiency, i.e. $\eta=1$, (a) $E_t$ grows steadily with $G_s$ and $\kappa_m$. When $\eta=0.69$ (b), it plateaus at $2.2$ for $\kappa_m/\kappa_l=5$ and $G_s>20$. The color scales for (a) and (b) differ by close to a factor of $10$.}
		\label{fig:Rtheo}
	\end{figure}

	\section{EXPERIMENTALLY IMPROVED SIGNAL RECOVERY WITH SQUEEZING}
	\label{scanrate}
	
	The scan rate can be degraded not only by transmission losses, but also by distortions in the squeezed state \cite{malnou2018optimal, boutin2017effect} and added noise in the amplifier chain. In order to investigate how accurately Eq.\,\eqref{R} predicts the performance of a squeezing-enhanced haloscope, we construct an apparatus that mimics its microwave network, but without some of the cumbersome features. Specifically, in an axion haloscope the mechanically tunable cavity must reside in a large static magnetic field to enhance $\kappa_a$ and $n_A$. In our apparatus, the cavity has a fixed frequency $\omega_c  = 2\pi\times7.146$\,GHz and there is no applied field. Consequently our cavity, constructed from superconducting aluminum, has much lower intrinsic loss than a copper haloscope cavity. We create additional loss through an explicit port that extracts energy from the cavity at a rate $\kappa_l =2\pi \times\ 100$\,kHz, a typical value for a haloscope cavity. We introduce an axion-like signal into our cavity through a microwave generator connected to a second port weakly coupled at rate $\kappa_a = 2\pi \times 100$\,Hz. Finally, a third port couples the cavity mode to an SSR with a rate chosen to be close to the optimum value for the case with ($\kappa_m =10 \kappa_l$)  or without ($\kappa_m =1.5 \kappa_l$) squeezing, creating a physical realization of the model in Fig.\,\ref{fig:schematic}.  
	
\begin{figure}[!h]
		\centering
		\includegraphics[scale=0.49]{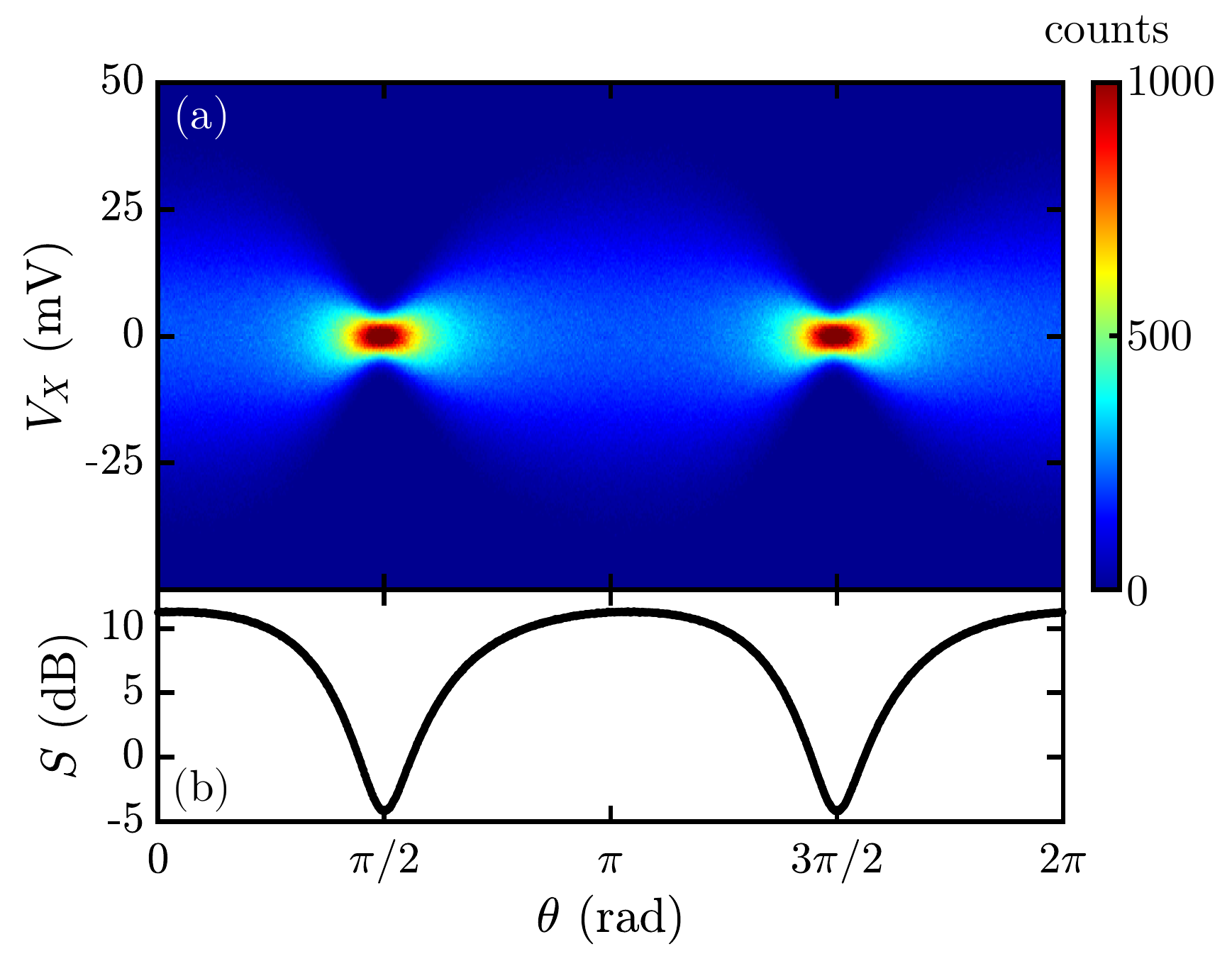}
		\caption{Histograms of voltage fluctuations $V_X$ (a) and corresponding vacuum squeezing (b) $S=\sigma_\mathrm{on}^2/\sigma_\mathrm{off}^2$ measured along $\hat{X}_{\mathrm{out},m}$, as a function of the SQ-AMP relative phase $\theta$.}
		\label{fig:squeezing}
	\end{figure}

	For the largest increase in scan rate, the SSR should be attached to the cavity measurement port with as little transmission loss as possible. To investigate the transmission loss independent of the cavity loss, we use the fact that the JPAs are narrow band ($\sim 5$ MHz), tunable amplifiers and detune both the SQ and AMP far off cavity resonance ($\sim 10$ MHz) so that the squeezed state is promptly reflected from the cavity.  Figure~\ref{fig:squeezing} illustrates our ability to efficiently generate, transport, and amplify a squeezed state in this off-resonance configuration. It shows histograms of the measured voltage in the AMP's amplified quadrature $\hat{X}_{\mathrm{out},m}$ as a function of $\theta$, the phase between the amplified quadrature of the SQ and the amplified quadrature of the AMP. When $\theta=\pi/2$ or $3\pi/2$, one quadrature is first squeezed and then amplified. At these points, comparing the output noise variances $\sigma^2_\mathrm{on}$ and $\sigma^2_\mathrm{off}$ measured with SQ on and off (not pumped), respectively \cite{malnou2018optimal}, we obtain a squeezing $S=\sigma^2_\mathrm{on}/\sigma^2_\mathrm{off}=-4.5\pm0.1$\,dB. We emphasize that this is not an inferred squeezing; rather, we directly measure an overall $4.5$\,dB reduction in the noise floor of $\hat{X}_{\mathrm{out},m}$ over the whole bandwidth of the quadrature measurement. This amount of squeezing (consistent with our estimate of $\eta=0.69\pm0.01$, see Appendix~\ref{sup:losslines}) required particular care in reducing the transmission losses between the two JPAs, which was facilitated by using flux-pumped JPAs \cite{yamamoto2008flux,zhou2014high,pogorzalek2017hysteretic}.   
	
\begin{figure}[!h]
		\centering
		\includegraphics[scale=0.35]{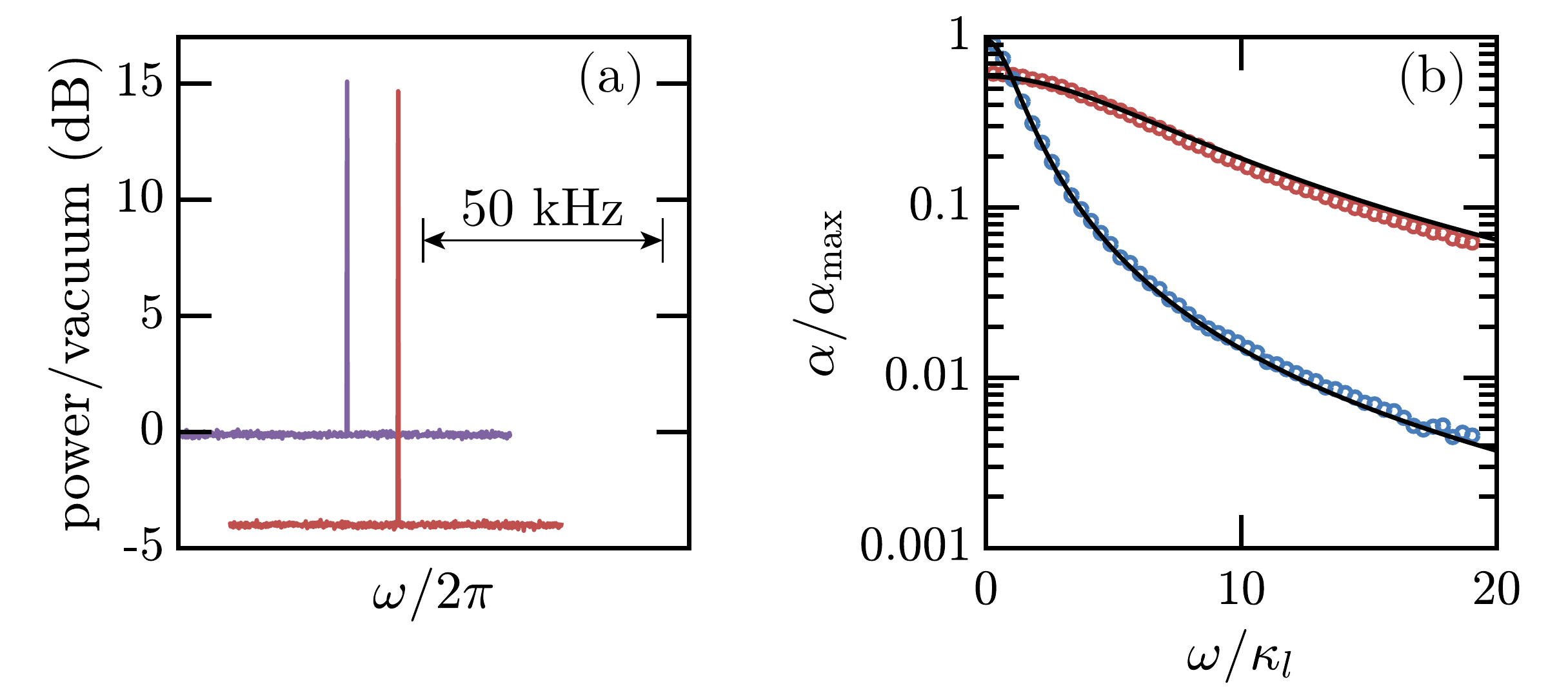}
		\caption{Improvement in microwave tone's signal visibility due to squeezing. Power spectra normalized to the unsqueezed vacuum power, with (red) and without (purple) squeezing are shown in (a) for a tone $1$\,MHz detuned from the resonant frequency of the overcoupled cavity ($\kappa_m=10\kappa_l$). The $x$-axis has been shifted between the two situations for visual clarity. In (b), the visibility $\alpha(\omega)$ is measured as a function of the tone's detuning from cavity resonance ($\omega=0$) for two cases: no squeezing while near critical coupling ($\kappa_m=1.5\kappa_l$, blue circles), and squeezing while strongly overcoupled ($\kappa_m=10\kappa_l$, red circles). In both cases $\alpha(\omega)$ is normalized to the expected maximum value $\alpha_\mathrm{max}=\alpha(0)$ evaluated at $\kappa_m=\kappa_l$. The theoretical expectation for each case is superimposed as a black curve, calculated using $\kappa_l=2\pi\times100$\,kHz, $\eta=0.69$ and $G_s=13$\,dB.}
		\label{fig:SNRimpr}
	\end{figure}

	This large amount of delivered squeezing implies that the SSR should improve our ability to resolve a weakly coupled signal detuned from cavity resonance. To demonstrate this improvement, we center the SQ and AMP bands on cavity resonance, overcouple the cavity's measurement port ($\kappa_m=10\kappa_l$), and set the phase $\theta$ to $\pi/2$ (see Fig.\,\ref{fig:squeezing}). We complete the single-quadrature measurement by mixing down $\hat{X}_{\mathrm{out},m}$ with a local oscillator (LO) at the cavity resonant frequency and computing its power spectral density. As for any such measurement, the frequency component at $\omega$ in the down-mixed output is a linear combination of two frequency components at the mixer's high-frequency input, $\pm\omega$ detuned from the LO frequency. Figure~\ref{fig:SNRimpr}a shows the spectral density when the tone is $1$\,MHz detuned from the cavity's resonance. Comparing two situations, SQ on and SQ off, the signal visibility improves by roughly $4$\,dB in the presence of squeezing, with $0.5$-dB degradation from cavity loss.
	
	To estimate the associated increase in scan rate from the SSR, we step the tone across the cavity's resonance and measure the visibility $\alpha(\omega)$ at each detuning $\omega$. In order to compare the optimal squeezed and unsqueezed cases, 
	we measure $\alpha(\omega)$ for two cases, displayed in Fig.\,\ref{fig:SNRimpr}b: for $\kappa_m=10\kappa_l$ with squeezing, and for $\kappa_m=1.5\kappa_l$ without squeezing (see Appendix~\ref{sup:dual} for two other complementary cases). Without squeezing, $\alpha(0)$ is greater but the bandwidth is poor. With squeezing, $\alpha$ remains relatively high as $\omega$ increases. We extract $R\propto\int_{-\infty}^\infty\alpha(\omega)^2d\omega$ for the two cases, and obtain the estimated scan rate enhancement $E_e=2.05\pm0.04$. By independently measuring $\eta$, $\kappa_m$, and $G_s$, we calculate expected values for $\alpha(\omega)$, shown as solid lines, in excellent agreement with the measured values in Fig.\,\ref{fig:SNRimpr}b. Finally, from the expected $\alpha(\omega)$ we calculate $E_t=2.11\pm0.07$, also in quantitative agreement with the data-based estimate $E_e$.

	\section{SQUEEZING-ENHANCED SEARCH FOR AN AXION-LIKE SIGNAL} \label{sec:faxion detection}
		
	In a real axion search, the aim is to detect a signal many orders of magnitude smaller than that in Fig.\,\ref{fig:SNRimpr}. Inferring the presence or absence of an axion at each frequency requires combining the measured powers from many adjacent cavity tunings \cite{brubaker2017first,du2018search}. Furthermore, over a long integration time the benefit inferred from Fig.\,\ref{fig:SNRimpr} is vulnerable to practical non-idealities. Drifts of either JPA's gain, drifts of the SQ-AMP relative phase $\theta$, non-Gaussian noise processes, and interfering rf or IF tones are of particular concern. In this section, we demonstrate that our SSR indeed matches the performance presented in the previous section when searching for a feeble tone over a wide frequency range.

\begin{figure}[!h]
	    \centering
    	\includegraphics[scale=0.49]{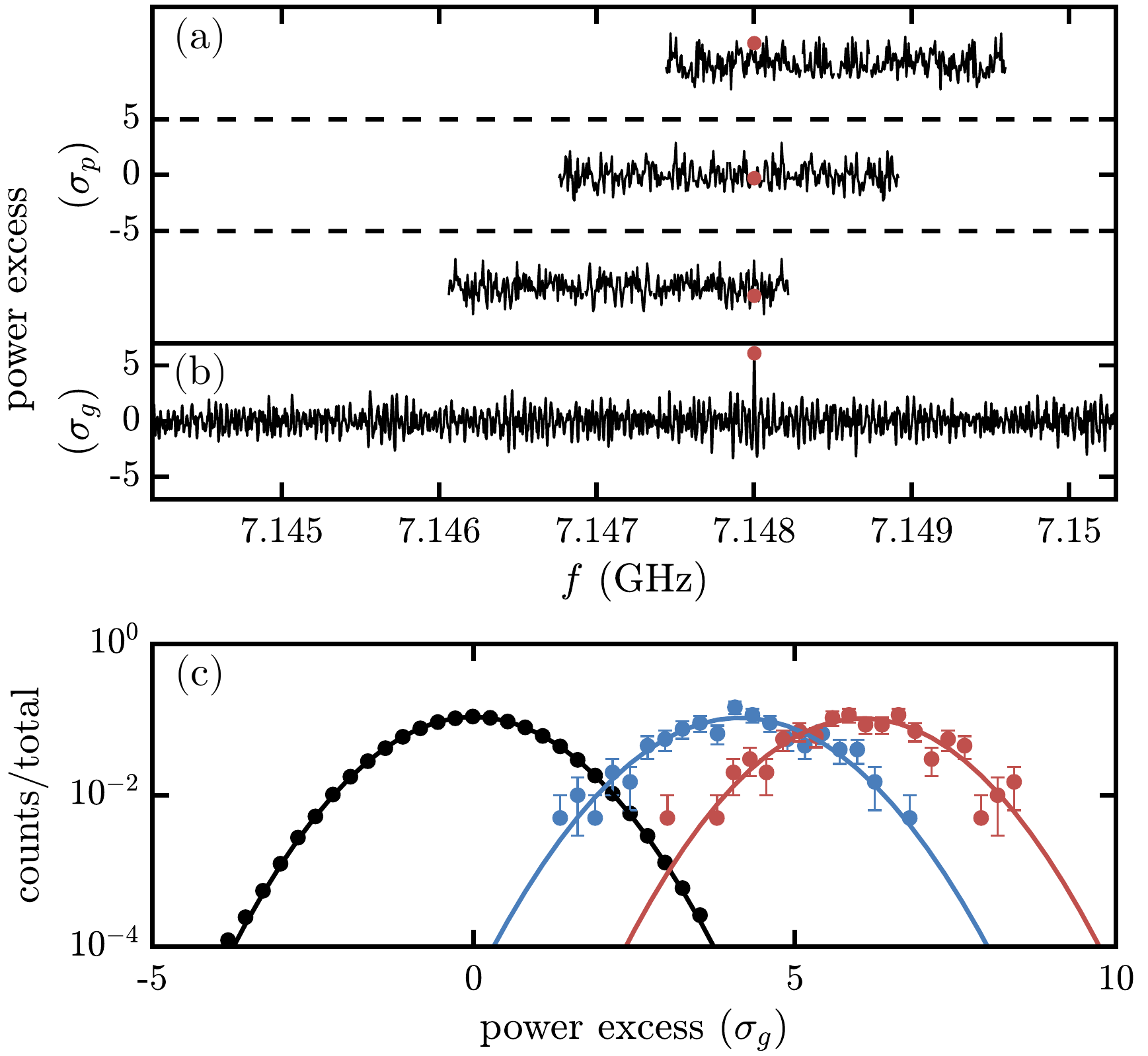}
	    \caption{Response of the SSR in the presence of a faxion. The acquired power spectra are symmetrized (a), rescaled and then shifted in frequency to align all the frequency bins containing the faxion with each other (red dots), thereby effectively tuning the cavity. The power excess of the processed spectra is plotted in units of their standard deviation $\sigma_p$. Combining all the spectra into a grand spectrum (b), whose excess power is plotted in units of its standard deviation $\sigma_g$, a large power excess is observed at the initial frequency of the faxion. Repeating this measurement allows us to compute distributions of faxion powers (c). When squeezing with $\kappa_m=10\kappa_l$, the powers (red points) are drawn from a Gaussian distribution $\mathcal{N}(\mu_s,\sigma_g^2)$ (solid line), with $\mu_s=6.05\pm0.07$. When not squeezing with $\kappa_m=1.5\kappa_l$, the powers (blue points) are drawn from $\mathcal{N}(\mu_u,\sigma_g^2)$ (solid line), with $\mu_u=4.15\pm0.07$. The powers (black points) drawn from the faxion-less noise distribution $\mathcal{N}(0,\sigma_g^2)$ (solid line) are also represented for comparison.}
		\label{fig:faxionsearch}
	\end{figure}

	We attempt to detect a fake axion, or ``faxion," sent through the cavity weakly coupled port. It is synthesized from a randomly modulated microwave tone, whose power is adjusted such that the faxion spectral density emerging from the cavity is roughly $1\%$ of vacuum, and whose width is broadened to roughly $\Delta_a\approx9$\,kHz, comparable to expectations for a realistic axion at frequency $\omega_a \approx 2\pi\times7$\,GHz \cite{brubaker2017haystac}. Stepping the faxion tone frequency ``backwards" past a stationary cavity simulates a realistic axion search without the hardware demands imposed by a tunable cavity. The faxion's initial frequency is chosen randomly within a $2$-MHz window around the cavity resonance, and is then tuned in discrete $-10$-kHz steps over a $4$-MHz window. At each tuning step, we record an output power spectral density as described in Sec.\,\ref{scanrate}. These spectra are mixed down and referred to the mixer's input by symmetrizing about $\omega=0$. Spectra are artificially shifted in steps of $+10$\,kHz to simulate cavity tuning. We then rescale the spectra by $\alpha(\omega)$ such that frequency bins with higher sensitivity to the faxion are weighted more. These rescaled, shifted spectra are then added into a grand spectrum  (see Appendix~\ref{sup:processing} and Ref.\,\cite{brubaker2017haystac}). With this procedure, the faxion's contributions in each individual spectrum add at its initial frequency as if we had tuned the cavity, creating a clear excess of power in the grand spectrum. Figure~\ref{fig:faxionsearch}a presents some symmetrized spectra, obtained while squeezing with an overcoupled cavity ($\kappa_m = 10\kappa_l$), in which the faxion excess power is too small to rise above vacuum fluctuations. The spectra were obtained at different fictitious cavity tunings, normalized to their measured standard deviations, and vertically offset from one another for visual clarity. In the resulting grand spectrum (Fig.\,\ref{fig:faxionsearch}b), a prominent faxion peak emerges with $6\sigma$ visibility. 
	
	In order to extract a scan rate enhancement from such a realistic signal search, we acquire two distributions of faxion powers, one when squeezing and one when not squeezing, with near-optimal $\kappa_m$ for each case. We obtain each distribution by repeating the faxion injection and detection protocol $200$ times. Over the course of the measurement, which takes roughly nine hours per configuration, we keep the relative phase $\theta$ between SQ and AMP quadratures at $\pi/2$ via a feedback loop (see Appendix~\ref{sup:PLL}). Figure~\ref{fig:faxionsearch}c displays the two faxion power distributions, as well as the noise power distribution. The no-faxion distribution of mean zero and variance $\sigma_g^2$ is guaranteed by the central limit theorem to be Gaussian distributed, $\mathcal{N}(0,\sigma_g^2)$. The faxion adds a small mean excess of power $\mu$ insufficient to enlarge the variance. The faxion distribution is therefore $\mathcal{N}(\mu,\sigma_g^2)$. The signal and noise distributions separate as the total measurement time squared, and thus the speed-up due to squeezing is equal to $E_m = (\mu_s/\mu_u)^2$, where $\mu_s$ $(\mu_u)$ is the mean of the faxion power distribution obtained when squeezing (not squeezing). The two distributions of faxion powers are characterized by $\mu_s=6.05\pm0.07$ and $\mu_u=4.15\pm0.07$, leading to a measured scan rate enhancement of $E_m=2.12\pm0.08$, in quantitative agreement with the estimate $E_e$ obtained from visibility measurements in Sec.\,\ref{scanrate}. 	
	

	\section{CONCLUSION} 
	
		
	The low signal-to-noise ratios achievable in state-of-the-art axion haloscopes \cite{du2018search, zhong2018results, chung2016cultask, mcallister2017organ} and similar resonant searches for faint electromagnetic signals sourced by dark matter axions \cite{ouellet2018first, silvafeaver2017design, majorovits2017madmax} make the timescales over which they can sweep out appreciable fractions of parameter space unreasonably long. Partially replacing the vacuum noise in an axion haloscope with squeezed vacuum circumvents the standard quantum limit on the noise of the measurement apparatus, enabling high sensitivity over a broader bandwidth and a more rapid dark matter search. Squeezing moreover moves haloscopes into a qualitatively distinct design parameter space: whereas a quantum limited haloscope's scan rate plateaus with improving microwave transmission efficiencies $\eta$, a sub-quantum limited haloscope will benefit almost arbitrarily as microwave losses are further reduced. Thus, haloscopes can reap larger benefits from increased efficiency (for example, a tenfold scan rate enhancement at $\eta=0.91$) as low-loss quantum technologies such as on-chip circulators and directional amplifiers mature \cite{sliwa2015reconfigurable, macklin2015a,chapman2017widely}.

	
	

	\section*{Acknowledgements} 
	
	We thank Kyle Thatcher for his help in the design and fabrication of the SSR mechanical parts and Felix Vietmeyer for his help in the design and fabrication of room temperature electronics. This work was supported by the National Science Foundation, under Grants No.\,PHY-1607223 and No.\,PHY-1734006, and by the Heising-Simons Foundation under Grant No.\,2014-183.
	
	\appendix
	
	
	\section{Theory of the SSR operation}
	\label{sup:SSR}
	
	In this appendix, we track the propagating electromagnetic fields through the SSR and cavity, so as to derive the susceptibility matrix of the entire system. We then calculate the signal visibility $\alpha(\omega)$ and the scan rate $R$. We first consider the case where the propagating fields experience no loss (the cavity mode still decays partially out its loss port), which nonetheless faithfully illustrates the utility of the SSR, then we treat the full system in the presence of transmission losses.
		
    \subsection{Lossless case}
    \label{sup:SSRnoloss}
    
	We model the energy exchange between the cavity's ports in its own rotating frame. The time evolution of the cavity is governed by the Heisenberg-Langevin equation
	\begin{equation}
	\label{EOM}
	\frac{d\hat{A}}{dt} = -\frac{\kappa_T}{2} \hat{A}(t) + \sum_{j} \sqrt{\kappa_j} \hat{a}_{\mathrm{in},j}(t),
	\end{equation}
	where $\hat{A}$ is the cavity ladder operator, $\kappa_T=\kappa_m+\kappa_l+\kappa_a$, and $\hat{a}_{\mathrm{in},j}$ ($j=m,l,a$) are the annihilation field operators of the input modes incident on ports indexed by $m,$ $l$, and $a$. At the measurement port, we physically separate input ($\hat{a}_{\mathrm{in},m}$) and output ($\hat{a}_{\mathrm{out},m}$) fields with a circulator. Given the input-output relations $\hat{a}_{\mathrm{out},j}(t) = \hat{a}_{\mathrm{in},j}(t) - \sqrt{\kappa_j}\hat{A}(t)$, the input and output field operators are related in the Fourier domain by
	\begin{equation}
	\label{IOcav}
	  \hat{a}_{\mathrm{out},j}(\omega)=\sum_k\chi_{jk}(\omega)\hat{a}_{\mathrm{in},k}(\omega),
	\end{equation}
	where
	\begin{equation}
	\label{suceptibility}
	    \chi_{jk}(\omega)=\frac{-\sqrt{\kappa_j\kappa_k} + (\kappa_T/2+i\omega)\delta_{jk}}{\kappa_T/2+i\omega}
	\end{equation}
	are the elements of a $3\times3$ susceptibility matrix fully describing the behavior of the cavity \cite{clerk2010introduction}.
	
	By cascading the input-output relations for each element presented in Fig.\,\ref{fig:schematic}, we can calculate the benefit from squeezing. We work in the quadrature basis and consider the vector of input quadratures $\vec{x}_\mathrm{in}=[\hat{X}_{\mathrm{in},m},\hat{X}_{\mathrm{in},l},\hat{X}_{\mathrm{in},a}]^\mathrm{T}$ aligned with our squeezing. We will calculate the SSR/cavity susceptibility matrix $\boldsymbol{\Xi_{X}}$, in terms of which the vector of output quadratures is $\vec{x}_\mathrm{out}=\boldsymbol{\Xi_{X}}\vec{x}_\mathrm{in}$. 
	
	The first element in the system is the SQ, which performs a one-mode squeezing (OMS) operation on the measurement port's input quadrature:
	\begin{equation} \label{IOSQ}
	\vec{x}_s = \boldsymbol{S_X} \vec{x}_\mathrm{in} = \begin{bmatrix} \frac{1}{\sqrt{G_s}} && 0 && 0\\
	                                0 && 1 && 0 \\
	                                0 && 0 && 1 \\\end{bmatrix} \vec{x}_\mathrm{in},
	\end{equation}
    where the subscript $s$ refers to the SQ output port and $G_s$ is the SQ single-quadrature power gain. The OMS operation also amplifies the other quadrature of the measurement port mode, $\hat{Y}_{\mathrm{in},m}$ by $\sqrt{G_s}$ in order to preserve the Heisenberg uncertainty relation. We do not track the evolution of the $\hat{Y}$ quadrature here, as it is irrelevant for SSR performance.
	
	Next, the cavity transforms the quadrature operators. In the cavity rotating frame, the vectors of quadrature operators are obtained from the vectors of ladder operators by
	\begin{equation} \label{quadladder}
    \begin{bmatrix} \vec{x}\\\vec{y} \end{bmatrix} = \frac{1}{\sqrt{2}} \begin{bmatrix} \boldsymbol{I_3} && \boldsymbol{I_3} \\ -i\boldsymbol{I_3} && i\boldsymbol{I_3}\end{bmatrix} \begin{bmatrix} \vec{a}(\omega)\\\vec{a}^\dagger(-\omega) \end{bmatrix},  
	\end{equation}
	or $[\vec{x},\vec{y}]^\mathrm{T}=\boldsymbol{P}[\vec{a}(\omega),\vec{a}^\dagger(-\omega)]^\mathrm{T}$. Here $\boldsymbol{I_3}$ is the $3\times3$ identity matrix, and the mode ordering in each vector is $m,l,a$: for example, $\vec{a}(\omega) = [\hat{a}_m(\omega),\hat{a}_l(\omega),\hat{a}_a(\omega)]^\mathrm{T}$. Thus, the cavity susceptibility matrix in the quadrature basis $\boldsymbol{\Tilde{\chi}}$ is:	
	\begin{equation} \label{quadsusceptibility}
    \boldsymbol{\Tilde{\chi}} = \boldsymbol{P} \begin{bmatrix} \boldsymbol{\chi}(\omega) && \boldsymbol{0_3} \\ \boldsymbol{0_3} && \boldsymbol{\chi}^*(-\omega)\end{bmatrix} \boldsymbol{P}^{-1} =  \begin{bmatrix} \boldsymbol{\chi}(\omega) && \boldsymbol{0_3} \\ \boldsymbol{0_3} && \boldsymbol{\chi}(\omega)\end{bmatrix},
	\end{equation}
	where $\boldsymbol{0_3}$  is the null matrix. We see that the cavity's effect on field and quadrature operators is identical, which leads to
	\begin{equation} \label{IOcavSSR}
        \vec{x}_o = \boldsymbol{\chi}(\omega) \vec{x}_s,
	\end{equation}
	where $o$ refers to the cavity output port. 	
	
	Finally the AMP performs a second OMS operation so as to amplify the quadrature that the SQ originally squeezed:
	\begin{equation} \label{IOAMP}
	\vec{x}_\mathrm{out} = \boldsymbol{A_X} \vec{x}_o = \begin{bmatrix} \sqrt{G_a} && 0 && 0\\
	                                0 && 1 && 0 \\
	                                0 && 0 && 1 \\\end{bmatrix} \vec{x}_o,
	\end{equation}
    where $G_a$ is the AMP power gain. 
	
	The SSR and cavity thus transform input to output quadratures according to
	\begin{equation} \label{IOSSR}
        \vec{x}_\mathrm{out} = \boldsymbol{A_X} \boldsymbol{\chi}(\omega) \boldsymbol{S_X} \vec{x}_\mathrm{in} = \boldsymbol{\Xi_{X}} \vec{x}_\mathrm{in}.    
	\end{equation}

	We can then calculate the single-quadrature output spectral density matrix $\boldsymbol{\Sigma_{\mathrm{out,}X}} =\langle [\vec{x}_\mathrm{out}]^\dagger [\vec{x}_\mathrm{out}]^\mathrm{T} \rangle / 2\pi$, where the Hermitian conjugation here does not transpose the vector, in order to determine the total output signal and noise powers. Substituting for $\vec{x}_\mathrm{out}$ yields
	\begin{equation} \label{Sout}
    \boldsymbol{\Sigma_{\mathrm{out,}X}} = \boldsymbol{\Xi_{X}}^* \boldsymbol{\Sigma_{\mathrm{in,}X}} \boldsymbol{\Xi_{X}}^\mathrm{T},
	\end{equation}
	where $\boldsymbol{\Sigma_{\mathrm{in,}X}} =\langle [\vec{x}_\mathrm{in}]^\dagger [\vec{x}_\mathrm{in}]^\mathrm{T}\rangle / 2\pi$ is the input noise spectral density matrix. Since the SSR is connected to three incoherent and uncorrelated modes (see Fig.\,\ref{fig:schematic}), $\langle \hat{X}_{\mathrm{in},j} \hat{X}_{\mathrm{in},k} \rangle/2\pi = \delta_{jk}(n_{\mathrm{in,}j} + 1/2)$, where $n_{\mathrm{in,}j}$ is the input mean photon number per unit time per unit bandwidth (henceforth simply mean photon number) of mode~$j$. Thus,
	\begin{equation} \label{Sin}
    \boldsymbol{\Sigma_{\mathrm{in,}X}} = \begin{bmatrix} n_T+\frac{1}{2} && 0 && 0\\
	                                0 && n_T+\frac{1}{2} && 0 \\
	                                0 && 0 && n_A+\frac{1}{2} \\\end{bmatrix}.
	\end{equation}	
	 Taking the first entry in the matrix $\boldsymbol{\Sigma_{\mathrm{out,}X}}$ we obtain the single-quadrature output spectral density at the measurement port:
	\begin{equation}\label{SoutXm}
	\begin{aligned}
    \Sigma_{\mathrm{out,}X,m} \approx \frac{G_a}{B(\omega)} \bigg[ & n_A\kappa_a\kappa_m \\
    + & \left(n_T + \frac{1}{2}\right) \left(\kappa_l\kappa_m + \frac{\beta(\omega)}{G_s}\right) \bigg],
    \end{aligned}
	\end{equation}	
	where $B(\omega) = (\kappa_m + \kappa_l)^2/4 + \omega^2$, $\beta(\omega) = (\kappa_m - \kappa_l)^2/4 + \omega^2$, and we used $\kappa_a \ll \{\kappa_l,\kappa_m\}$ and $n_A \gg 1/2$.
	
	The visibility $\alpha(\omega)$, defined as the ratio of signal and noise spectral densities at the measurement port output, can be directly extracted from $\Sigma_{\mathrm{out,}X,m}$, since the signal is the term proportional to $n_A$ and the noise is $\Sigma_{\mathrm{out,}X,m}(n_A=0)$:
	\begin{equation}\label{eq:SNRapp}
    \alpha(\omega) \approx
    \frac{n_A\kappa_a\kappa_m}{\left(n_T + \frac{1}{2}\right)\left(\kappa_l\kappa_m + \frac{\beta(\omega)}{G_s}\right)}, 
	\end{equation}	
	which is identical to Eq.\,\eqref{SNR}. 
	
    We now consider a search protocol comprising many measurements of $\Sigma_{\mathrm{out,}X,m}$, as the cavity frequency is tuned by discrete steps $\delta_c$. The spectral scan rate $R$ is obtained by taking the limit:
    \begin{equation}
	\label{Rdef}    
	R = \lim_{\substack{\delta_c\to 0 \\ \tau\to 0}} \frac{\delta_c}{\tau},
	\end{equation}
	where $\tau$ is the duration of each measurement. In the rest of this section we derive Eq.\,\eqref{R} for the scan rate. We begin by defining the signal-to-noise ratio (SNR) at a single tuning step as
    \begin{equation}
	\label{SNRdef}    
	\overline{\alpha}(\omega) = \sqrt{\tau\Delta_a}\frac{\alpha(\omega)}{2},
	\end{equation}
    which scales as the square root of the number of independent measurements of the power contained in the bandwidth $\Delta_a$, centered on a frequency detuned from cavity resonance by $\omega/2\pi$ \cite{dicke1946the,brubaker2017first}; the factor of 2 is a consequence of our single-quadrature measurement scheme (see Appendix~\ref{sup:singledouble}). 

    For $\delta_c\lesssim \kappa_T/2\pi$, multiple tuning steps will contribute to the SNR at each frequency. Without loss of generality, we evaluate the net SNR for some putative signal frequency $\omega_a$ which coincides exactly with the cavity resonance at a particular step. We then consider the contributions to the SNR of the surrounding $2n+1$ tuning steps, where $2\pi\delta_c=2\kappa_T/(2n+1)$. Contributions to the SNR add in quadrature, so the net squared SNR is 
    \begin{equation}
	\label{alpha_In}    
	\overline{\alpha}_{2n+1}^2 = \frac{\tau\Delta_a}{4}\cfrac{1}{\delta_c}\sum_{k=-n}^{k=n}\alpha^2(k2\pi\delta_c)\delta_c.
	\end{equation}
	In the limit $n\to\infty$, $\tau\to0$, we obtain the integrated squared SNR
	\begin{equation}
	\label{alpha_I}
	\overline{\alpha}_{I}^2 = \frac{\Delta_a}{4}\frac{1}{R} \int_{-\infty}^\infty\alpha^2(\omega) \frac{d\omega}{2\pi},
	\end{equation}
	where we have extended the limits of integration to $\pm\infty$: the contributions from tuning steps where the cavity is detuned by more than $\kappa_T$ are negligible.
	
	Note that our arbitrary choice of $\omega_a$ appears nowhere explicitly on the RHS of Eq.\,\eqref{alpha_I}; thus the integrated SNR will be frequency independent, insofar as the quantities contributing to $\alpha(\omega)$ remain constant as we tune the cavity. In practice, we set a certain target value of $\overline{\alpha}_I$ for which a real signal would appear as a sufficiently prominent peak in the grand spectrum (see e.g., Fig.\,\ref{fig:faxionsearch}b); this choice determines the required spectral scan rate $R$. Solving for $R$, we obtain	
	\begin{equation}
	\label{Rfull}    
	\begin{aligned}
	R &= \frac{\Delta_a}{4\overline{\alpha}_I^2} \int_{-\infty}^\infty\alpha^2(\omega) \frac{d\omega}{2\pi}\\
	&= \frac{\Delta_a\sqrt{G_s}n_A^2\kappa_a^2\kappa_m^2}{16\overline{\alpha}_I^2\left(n_T+\frac{1}{2}\right)^2\left[\kappa_l\kappa_m + \frac{1}{G_s}\left(\frac{\kappa_l-\kappa_m}{2}\right)^2\right]^{3/2}},
	\end{aligned}
	\end{equation}
	in agreement with Eq.\,\eqref{R}. In a comparison of squeezed and unsqueezed scan rates, the $\Delta_a/4\overline{\alpha}_I^2$ factor drops out; thus, we need only compare the integral in Eq.\,\eqref{Rfull} for the two cases.
	
	\subsection{SSR operation in the presence of transmission losses}
	\label{sup:SSRwloss}
	
    Here  we generalize the calculations of Appendix~\ref{sup:SSRnoloss} to account for imperfect power transmission efficiency $\eta$ between SQ and AMP. Integrating the squared SNR over all frequencies, we obtain the scan rate as a function of $\eta$, used in Fig.\,\ref{fig:Rtheo}. For simplicity, we treat the transmission efficiency $\lambda$ between the SQ and the cavity as identical to that between the cavity and the AMP, hence $\eta = \lambda^2$. Note that, in principle, loss between SQ and cavity is slightly less harmful than loss between cavity and AMP, as loss that occurs after the cavity degrades the signal along with the squeezing. 
		
    As in Appendix~\ref{sup:SSRnoloss}, we track the vector of input quadratures $\vec{x}_\mathrm{in}$ throughout the SSR and cavity system, neglecting the orthogonal quadratures. First, the SQ performs the OMS operation defined by Eq.\,\eqref{IOSQ}: $\vec{x}_s=\boldsymbol{S_X} \vec{x}_\mathrm{in}$. Next we must account for transmission losses between the SQ and the cavity, modeled as a beamsplitter interaction between the quadrature operators of the measurement line at the SQ output $\{\hat{X}_{s,m},\hat{Y}_{s,m}\}$, and those of another, uncontrolled mode $\{\hat{X}_{s,\lambda},\hat{Y}_{s,\lambda}\}$. We could define generalized quadrature vectors $\vec{x}_s$ and $\vec{y}_s$ to include $\hat{X}_{s,\lambda}$ and $\hat{Y}_{s,\lambda}$ respectively, but we do not need to keep track of the electromagnetic field's evolution in loss modes. Among the modes we do keep track of, only the measurement port experiences the loss:
    \begin{equation} \label{IOloss}
	\vec{x}_{i} = \begin{bmatrix} \sqrt{\lambda} && 0 && 0\\
	                                0 && 1 && 0 \\
	                                0 && 0 && 1 \\\end{bmatrix} \vec{x}_s,
	\end{equation}
	or $\vec{x}_{i} = \boldsymbol{L_X} \vec{x}_s$. Here $i$ refers to the cavity's input, and $0\leq\lambda\leq1$ is the single-sided transmission efficiency. 
	
	Since we are not tracking the SQ-to-cavity propagating loss mode, the unsqueezed vacuum that it introduces enters as an added noise term,
	\begin{equation} \label{noise}
    \boldsymbol{N_X} = \begin{bmatrix} (n_T+\frac{1}{2})(1-\lambda) && 0 && 0\\
	                                0 && 0 && 0 \\
	                                0 && 0 && 0 \\\end{bmatrix}.
	\end{equation}
	We thus have the single-quadrature noise spectral density at the cavity input: $\boldsymbol{\Sigma_{i,X}} = [\boldsymbol{L_X} \boldsymbol{S_X}]^*\boldsymbol{\Sigma_{\mathrm{in,}X}}[\boldsymbol{L_X} \boldsymbol{S_X}]^\mathrm{T} + \boldsymbol{N_X}$, where $\boldsymbol{\Sigma_{\mathrm{in,}X}}$ is given by Eq.\,\eqref{Sin}. This expression yields
	\begin{equation} \label{Sbeta1}
    \boldsymbol{\Sigma_{i,X}} = \begin{bmatrix} \left(n_T+\frac{1}{2}\right)\left(\frac{\lambda}{G_s}+1-\lambda\right) && 0 && 0 \\ 
    0 && n_T+\frac{1}{2} && 0 \\
    0 && 0 && n_A+\frac{1}{2} \\\end{bmatrix},
	\end{equation}
	where both the partially attenuated squeezed vacuum term ($\propto\lambda/G_s$), and the unsqueezed contribution from the loss mode ($\propto 1-\lambda$) appear clearly. Note that in the absence of loss, $\boldsymbol{\Sigma_{i,X}} = \boldsymbol{S_X}^*\boldsymbol{\Sigma_{\mathrm{in,}X}}\boldsymbol{S_X}^\mathrm{T}$, in agreement with Eq.\,\eqref{Sout}.
	
    We can now calculate $\boldsymbol{\Sigma_{\mathrm{out,}X}}$, the output spectral density matrix along $\vec{x}$. Given Eq.\,\eqref{quadsusceptibility} for the cavity's susceptibility in the quadrature basis, we may write $\boldsymbol{\Sigma_{o,X}} = \boldsymbol{\chi}^*(\omega)\boldsymbol{\Sigma_{i,X}}\boldsymbol{\chi}(\omega)^\mathrm{T}$. Then losses between cavity and AMP are accounted for in the same manner as before, and finally the AMP performs another OMS operation, amplifying the SQ squeezed quadrature. We thus obtain
	\begin{equation} \label{Soutloss}
	\begin{aligned}
    \boldsymbol{\Sigma_{\mathrm{out,}X}} & = \boldsymbol{A_X}^*[\boldsymbol{L_X}^*\boldsymbol{\Sigma_{o,X}}\boldsymbol{L_X}^\mathrm{T}+\boldsymbol{N_X}]\boldsymbol{A_X}^\mathrm{T}\\
    & =  \boldsymbol{A_X}^*[\boldsymbol{L_X}^*\boldsymbol{\chi}^*(\omega)\boldsymbol{\Sigma_{i,X}}\boldsymbol{\chi}(\omega)^\mathrm{T}\boldsymbol{L_X}^\mathrm{T}+\boldsymbol{N_X}]\boldsymbol{A_X}^\mathrm{T}
    \end{aligned}
    \end{equation}
    as the analog of Eq.\,\eqref{Sout} in the presence of loss.
    
    The first entry in the matrix $\boldsymbol{\Sigma_{\mathrm{out,}X}}$ is the output spectral density at the measurement port along $\hat{X}$:
    \begin{equation} \label{SoutXmloss}
    \begin{aligned}
    \Sigma_{\mathrm{out,}X,m} = & \left(n_T+\frac{1}{2}\right)G_a(1-\lambda)\\
    + \frac{G_a\lambda}{B(\omega)}\bigg[ & \left(n_A + \frac{1}{2}\right)\kappa_a\kappa_m\\
    + & \left(n_T + \frac{1}{2}\right) \left(\kappa_l\kappa_m + \left(1-\lambda+\frac{\lambda}{G_s}\right)\beta(\omega)\right)\bigg],
    \end{aligned}
    \end{equation}
    where $B(\omega)$ and $\beta(\omega)$ are defined as in Eq.\,\eqref{SoutXm}, and the same approximations have been made.
    
    From $\Sigma_{\mathrm{out,}X,m}$ we can extract the signal visibility $\alpha(\omega)$ in the presence of loss:
    \begin{equation} \label{SNRloss}
    \begin{aligned}
    \alpha(\omega) = & \frac{\lambda n_A}{\left(n_T + \frac{1}{2}\right)} \times \\
    & \frac{\kappa_a\kappa_m}
    {B(\omega)\left(1-\lambda\right) + \lambda\left[\kappa_l\kappa_m + \left(1-\lambda+\frac{\lambda}{G_s}\right)\beta(\omega)\right]}.
    \end{aligned}
	\end{equation}
    When $\lambda=1$, Eq.\,\eqref{SNRloss} reduces to Eq.\,\eqref{SNR}. 
    
    Finally, the scan rate enhancement $E_t$ in the presence of loss, presented in Fig.\,\ref{fig:Rtheo}b, is $\int_{-\infty}^\infty\alpha^2(\omega) d\omega$, normalized by the same integral with $G_s=1$ and $\kappa_m=2\kappa_l$.
	
	\section{Model for the axion field}
	\label{sup:axionmodel}
	Figure~\ref{fig:schematic} models the axion field as a fictitious generator that drives the cavity through a weakly coupled port. In this appendix, we relate the fictitious port coupling $\kappa_a$ and the power spectral density $n_A$ at the generator output to physical parameters normally found in the haloscope literature, and show that for representative values, the axion field acts as a classical force.
	
	The measurable axion-sourced power is obtained from the first term in the output spectral density, Eq.\,\eqref{SoutXm}. Referring to the cavity output and multiplying by the axion linewidth $\Delta_a$, the on-resonance ($\omega_a=\omega_c$) signal power is
	\begin{equation} \label{gen_model_pow}
	P_\mathrm{sig} = 4 \hbar \omega_a n_A \Delta_a \frac{\kappa_a \kappa_m}{(\kappa_m + \kappa_l)^2}.
	\end{equation}	
	The two model parameters $\kappa_a$ and $n_A$ can readily be related to physical parameters by comparing Eq.\,\eqref{gen_model_pow}, to e.g. Eq.\,(1) of Ref.\,\cite{alkenany2017design} evaluated on resonance. In our notation this expression takes the form
    \begin{equation}\label{eq:haloscope2}
    P_\text{sig} = \left(g_{a\gamma\gamma}^2\frac{\hbar c^3\rho_a}{\mu_0\omega_a^2}\right)\times\left(B_0^2VC_{mn\ell}\frac{\omega_c^2\kappa_m}{(\kappa_m + \kappa_l)^2}\right),
    \end{equation} 
	where, in the first set of parentheses, $g_{a\gamma\gamma}$ parametrizes the axion field's coupling to electromagnetism, $\rho_a$ is the local dark matter energy density, $c$ is the speed of light, $\mu_0$ is the vacuum permeability, and we have related the axion rest mass $m_a$ to the frequency of axion-induced photons $\omega_a$ as $m_ac^2 = \hbar \omega_a$. The second parenthetical expression contains properties of the haloscope: $B_0$ is the static magnetic field, $C_{mnl}$ is the cavity mode-dependent form factor, and $V$ is the cavity volume. Equating Eqs.\,\eqref{gen_model_pow} and \eqref{eq:haloscope2}, we obtain
	\begin{equation} \label{n_A times kappa_a}
	    n_A \kappa_a = \frac{g_{a\gamma\gamma}^2 \rho_a c^3}{4 \omega_a \mu_0 \Delta_a}B_0^2 C_{mnl} V.
	\end{equation}
    
    To derive a second expression relating $\kappa_a$ to $n_A$, we observe that the fictitious generator in our haloscope model may equivalently be represented as a second harmonic oscillator mode with very high occupancy but very weak coupling to the haloscope cavity. The axion field in any laboratory-scale volume constitutes such an oscillator mode, with resonant frequency $\omega_a$. Specifically, we model the oscillations of the axion field as a fictitious cavity occupying the same volume as the real haloscope cavity. The quanta of this fictitious cavity are axion particles, so its total occupancy is $N_A = V\rho_a/\hbar\omega_a$; it is  coupled to the haloscope cavity with interaction Hamiltonian $\hat{H}_\mathrm{int} = \hbar g (\hat{A} + \hat{A}^\dagger)(\hat{B} + \hat{B}^\dagger)$, where $g$ is the interaction strength and $\hat A$ ($\hat B$) is the annihilation operator of the haloscope (fictitious) cavity. The Hamiltonian of the closed system is $\hat{H} = \hat{H}_0 + \hat{H}_\mathrm{int}$, where $\hat{H}_0 = \hbar \omega_c(\hat{A}^\dagger \hat{A} + 1/2) + \hbar \omega_a(\hat{B}^\dagger \hat{B} + 1/2)$. Coupling the haloscope cavity to measurement and loss ports at rates $\kappa_m$ and $\kappa_l$, respectively, we write down the Heisenberg-Langevin equations of motion for the open system: 
    \begin{align}
    \frac{d\hat{A}}{dt} &= -i\omega_c \hat{A}(t) -i g [\hat{B}(t) + \hat{B}^\dagger(t)] - \frac{\kappa_m + \kappa_l}{2} \hat{A} \label{dAdt}  \\ 
     & + \sqrt{\kappa_m} \hat{a}_{\mathrm{in},m}(t) + \sqrt{\kappa_l} \hat{a}_{\mathrm{in},l}(t)  \nonumber \\
    \frac{d\hat{B}}{dt} &= -i\omega_a \hat{B}(t) -i g [\hat{A}(t) + \hat{A}^\dagger(t)], \label{dBdt}
    \end{align} 
    where we describe the input bath associated with the measurement (loss) port with the annihilation operator $\hat{a}_{\mathrm{in},m}$ ($\hat{a}_{\mathrm{in},l}$).
    
    We restrict ourselves to the classical limit of the system, with operators demoted to complex amplitudes, and the case where the resonances coincide, $\omega_a = \omega_c$, with no power entering via the measurement and loss ports, $a_{\mathrm{in},m} = a_{\mathrm{in},l} = 0$. Transforming into the rotating frame of the haloscope cavity, $\{A(t), B(t)\}\rightarrow \{A(t)e^{-i\omega_c t}, B(t)e^{-i \omega_c t}\}$ and making a rotating wave approximation, Eqs.\,\eqref{dAdt} and \eqref{dBdt} reduce to 
    \begin{align}
    \frac{dA}{dt} &= -i g B(t) -\frac{\kappa_m + \kappa_l}{2}A(t) \label{dAdt simplified}\\
    \frac{dB}{dt} &= -i g A(t) \label{dBdt simplified},
    \end{align}
    where $g$ plays the role of the field exchange rate corresponding to the power decay rate out of the axion cavity, $g = \kappa_a / 2$. These equations of motion describe an exchange of energy between the two cavities and a decay of that energy out from the haloscope cavity via the measurement and loss ports. We are interested in the steady-state ($dA/dt = 0$) output field $A_{\mathrm{out},m} = -\sqrt{\kappa_m} A$ when the occupancy of the axion cavity is $|B|^2 = N_A$. We find a steady-state occupancy of the haloscope cavity $|A|^2 = [\kappa_a / (\kappa_m + \kappa_l)]^2 N_A$, implying an output signal power of
    \begin{equation} \label{cav model pow}
    P_\mathrm{sig} =\hbar\omega_a |A_{\mathrm{out},m}|^2 = \frac{\hbar \omega_a N_A\kappa_a^2 \kappa_m}{(\kappa_m + \kappa_l)^2}.
    \end{equation}
    Equations~\eqref{gen_model_pow} and \eqref{cav model pow} for the output power must agree, implying 
    \begin{equation} \label{n_A over kappa_a}
    \frac{n_A}{\kappa_a} = \frac{N_A}{4 \Delta_a} = \frac{V\rho_a}{4\hbar\omega_a\Delta_a}. 
    \end{equation}
    In terms of physical parameters, Eqs.\,\eqref{n_A times kappa_a} and \eqref{n_A over kappa_a} yield
    \begin{align}
    n_A &= \frac{|g_{a\gamma\gamma}|\rho_a B_0 V}{4 \omega_a \Delta_a}\sqrt{\frac{C_{mnl} c^3}{\hbar \mu_0}}\\
    \kappa_a &= |g_{a\gamma\gamma}| B_0 \sqrt{\frac{C_{mnl} \hbar c^3}{\mu_0}}.
    \end{align} 
    
    \begin{table}[] 
        \centering
        \begin{tabular}{|l|l|} 
        \hline
        \textbf{quantity}& \textbf{value} \\ \hline
        $\rho_a$ & $0.45 \ \mathrm{GeV}/\mathrm{cm}^3$ \\ \hline
        $B_0$ & $9 \ \mathrm{T}$ \\ \hline
        $g_{a\gamma\gamma}$ & $-7.7 \times 10^{-24} \ \mathrm{eV}^{-1}$ \\ \hline
        $\Delta_a$ & $5 \ \mathrm{kHz}$ \\ \hline
        $\omega_a / 2\pi$ & $5 \ \mathrm{GHz}$ \\ \hline
        $V$ & $1.5 \ \mathrm{L}$ \\ \hline
        $C_{mnl}$ & $0.5$ \\ \hline
        \end{tabular} 
        \caption{Representative physical values used to estimate our model parameters.}
        \label{tab:phys vals}
    \end{table}

    Using representative values for the HAYSTAC experiment \cite{alkenany2017design} in the presence of a 5-GHz KSVZ \cite{kim1979weak, shifman1980can} axion shown in Table \ref{tab:phys vals}, we obtain the values for our model parameters shown in Table \ref{tab:param vals}. We see that the fictitious generator is well into the classical regime, $n_A\gg 1/2$, while its extremely feeble coupling $\kappa_a$ nonetheless makes its presence a challenge to detect. 
    \begin{table}[]
        \centering
        \begin{tabular}{|l|l|}
        \hline
        \textbf{parameter} & \textbf{value} \\ \hline
        $N_A$ & $3.3 \times 10^{16}$ \\ \hline
        $\kappa_a / 2\pi$ & $2.3\ \mathrm{\mu Hz}$ \\ \hline
        $n_A$ & $2.4 \times 10^7$ \\ \hline
        \end{tabular}
        \caption{Model parameter values calculated using the physical values from Table \ref{tab:phys vals}.}
        \label{tab:param vals}
    \end{table}

    \section{Single vs. double quadrature measurement}
    \label{sup:singledouble}
    
    Single-mode squeezing can only enhance sensitivity to displacements along a single quadrature of the cavity field. In a situation such as an axion search, the signal distributes its excess power equally between the two quadratures. Thus switching to single-quadrature measurement from double-quadrature measurement, currently the operational mode of choice of axion haloscopes \cite{brubaker2017first, du2018search}, seems detrimental. In this appendix, we show that in the absence of squeezing, neither measurement scheme has an advantage over the other.
    
    If we neglect amplifier added noise, the signal visibility $\alpha$ is independent of whether we measure one quadrature or both. Specifically, an axion signal characterized by its spectral density $S_a$ at the amplifier input divides itself equally as $S_a/2$ between the two quadratures. Similarly, the vacuum noise spectral density $\hbar\omega_a / 2$ divides its power equally as $\hbar\omega_a / 4$ between the quadratures. 
    
    However, we must account for the quantum limits on two-quadrature measurements. Any linear amplifier that measures both quadratures adds at least a second half-quantum of input-referred noise, evenly distributed between the two quadratures \cite{caves1982quantum}. An ideal double-quadrature measurement thus yields $\alpha_\mathrm{2Q} = S_a / \hbar \omega_a$ in each quadrature. By comparison, there is no quantum limit on single-quadrature amplification, so an ideal single-quadrature measurement yields $\alpha_\mathrm{1Q} = 2 S_a / \hbar \omega_a = 2\alpha_\mathrm{2Q}$ in the amplified quadrature.
    
    To make a fair comparison between the single-quadrature and double-quadrature cases, we must consider the improvement in the SNR (defined in Appendix~\ref{sup:SSRnoloss}). Because all pertinent differences between the two cases enter when considering a single tuning step, we neglect tuning in the following discussion. The SNR $\overline{\alpha}$ is given in terms of the spectral density ratio $\alpha$ by
    \begin{equation}
        \overline{\alpha}=\sqrt{\frac{N}{2}}\frac{\Delta_a}{\Delta}\alpha,
    \end{equation}
    where $\Delta_a/\Delta$ is the ratio of signal to noise bandwidths, and $N$ is the number of independent measurements of the voltage contained in the noise bandwidth $\Delta$. In considering the appropriate values of $N$ and $\Delta$ for the two cases of interest, we find two independent effects, each reducing $\overline{\alpha}_{\mathrm{1Q}}$ by a factor of $\sqrt{2}$ relative to $\overline{\alpha}_{\mathrm{2Q}}$. Together, these effects cancel the apparent benefit stemming from the absence of quantum noise limits in the single-quadrature case.
  
    First, the Nyquist theorem guarantees that there are $N_\mathrm{2Q}=2\tau\Delta_\mathrm{2Q}$ independent measurements of the noise voltage in a double-quadrature measurement of duration $\tau$ and bandwidth $\Delta_\mathrm{2Q}$, where the factor of 2 counts the two independent quadrature amplitudes for each resolved Fourier component. Thus there are  $N_\mathrm{1Q}=\tau\Delta_\mathrm{1Q}$ measurements of the noise voltage in a single-quadrature measurement of bandwidth $\Delta_\mathrm{1Q}$.

    Second, noiseless single-quadrature measurement with a parametric amplifier creates an irreversible ambiguity between the output signal and idler frequencies, equally spaced about the amplifier band center. This ambiguity necessitates mapping amplifier outputs at a given detuning from band center to the input signal and idler frequencies (see Appendix~\ref{sup:processing}). The consequent addition of spectral densities, half of which are guaranteed not to have an axion-induced excess power, effectively increases the noise bandwidth in a single-quadrature measurement by a factor of $2$ relative to the signal bandwidth: $\Delta_\mathrm{1Q}=2\Delta_a$. In the standard double-quadrature measurement scheme, the signal and noise bandwidths are equal: $\Delta_\mathrm{2Q}=\Delta_a$.
    
    Putting this all together, the two measurement schemes are seen to be equivalent: 
    \begin{equation}
    \begin{aligned}
    \overline{\alpha}_{\mathrm{1Q}} &= \sqrt{\frac{\tau\Delta_\mathrm{1Q}}{2}}\frac{\Delta_a}{\Delta_
    \mathrm{1Q}}\alpha_\mathrm{1Q} \\
    &= \sqrt{\frac{2\tau\Delta_\mathrm{2Q}}{2}}\frac{\Delta_a}{2\Delta_\mathrm{2Q}}2\alpha_\mathrm{2Q} \\
    &= \overline{\alpha}_{\mathrm{2Q}}.
    \end{aligned}
    \end{equation}
    The first line agrees with Eq.\,\eqref{SNRdef} for the single-quadrature SNR. 

	\section{Characterization of the experimental apparatus}
	In this appendix we discuss and characterize key features of the SSR, including its control electronics. As the performance of the SSR is primarily limited by its transmission losses, many of our design choices were driven by loss mitigation. Additionally, we designed the experiment to have a minimum number of adjustable controls, so that the faxion search could be easily automated.
	
	\subsection{Flux-pumped JPAs}
	\label{sup:fluxJPAs}	
	The requirement of running the SSR automatically for several days makes monochromatically current-pumped JPAs a poor choice for SQ and AMP, as the strong current pump tone remains present at the center of the amplification band output \cite{castellanos2008amplification,mallet2011quantum,malnou2018optimal}. If not canceled by a $\pi$-phase shifted tone of equal amplitude, the SQ pump would reach and saturate the AMP. Furthermore, with insufficient microwave isolation the AMP pump could also be reflected from the cavity and perturb the AMP operating point. Keeping both cancellation tones stable in amplitude and phase, while controlling both SQ and AMP pumps to obtain good gain and good squeezing is impractical over a long period of time.
	
\begin{figure}[!h]
	    \centering
    	\includegraphics[scale=0.49]{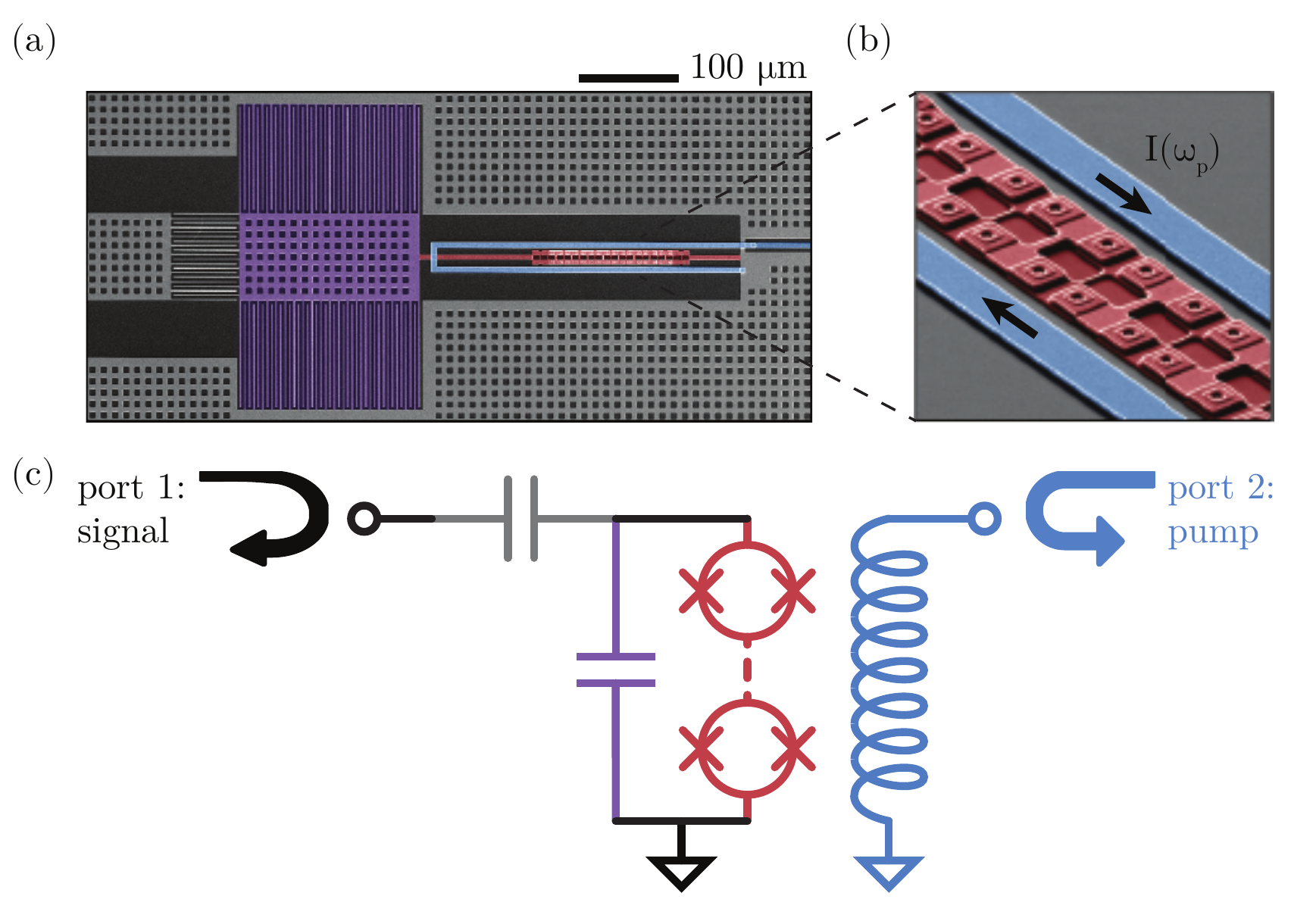}
	    \caption{A flux-pumped JPA, similar to those used in the SSR. It is fabricated in the standard niobium-aluminum-niobium process in a coplanar waveguide geometry. A scanning electron microscope image (a) shows, in false color, the main elements of the JPA: a $550$-fF interdigital capacitor (purple), an inductance comprising a SQUID array of $6$-\micro A Josephson junctions (red), and a flux line, shorted to ground (blue). A closer view (b) shows several SQUID loops (red), surrounded by the flux line (blue). The pump current $I(\omega_p)$ is circulating back and forth along the flux line, thereby favoring the differential mode current and attenuating the common-mode current in the SQUID loops. The JPA equivalent circuit (c) represents the two-port device: an incoming signal on port 1 is reflected and amplified. On port 2, the pump at $\omega_p=2\omega_0$ modulates the SQUID inductance.}
	    \label{fig:FPJPA}
	\end{figure}

	Flux-pumped JPAs do not require cancellation tones, because they are pumped at twice their operating frequency, far outside the bandwidth of the receiver chain. In addition, their topology is such that pump and signal propagate along spatially distinct paths, ensuring good isolation between the two (see Fig.\,\ref{fig:FPJPA}). This also negates the need for a directional coupler in front of the chip, and therefore reduces the insertion loss along the squeezed state path. When flux pumping, a JPA can be thought of as a linear, frequency-tunable $LC$ resonator, capacitively coupled to a transmission line. Josephson junctions arranged to form a series array of superconducting quantum interference devices (SQUIDs) provide the tunable inductance. The bare resonance $\omega_0$ of the $LC$ resonator is tunable via a dc magnetic flux. The pump generates an ac magnetic flux which modulates the SQUID inductance at $\omega_p = 2\omega_0$, generating parametric amplification.

    We optimized several constraints to ensure efficient flux pumping. First, we took care in the design to ensure identical coupling between the pump line and each of the SQUIDs that collectively constitute the tunable inductor. Uniform coupling guarantees that the JPA's dynamics are spatially homogeneous, resulting in improved power handling. Second, we placed the pump line close to the SQUID array, in order to minimize the pump power required to drive the $LC$ resonator. At the same time, the coupling of the $LC$ circuit to the pump line is kept much lower than the coupling to the line carrying the signal, in order to avoid losing part of the signal through the pump port: the transmission between the two ports is kept below $-20\,$dB. Finally, we shaped the flux line as a U around the SQUID array, as shown in Figs.\,\ref{fig:FPJPA}a and b. With this configuration, the pump couples to the differential mode current circulating inside each SQUID loop and is isolated from the common-mode current, unidirectional across the SQUID array.
	
	\subsection{Experimental setup}	
	\label{sup:setup}
\begin{figure*}
		\centering
		\includegraphics[scale=0.6]{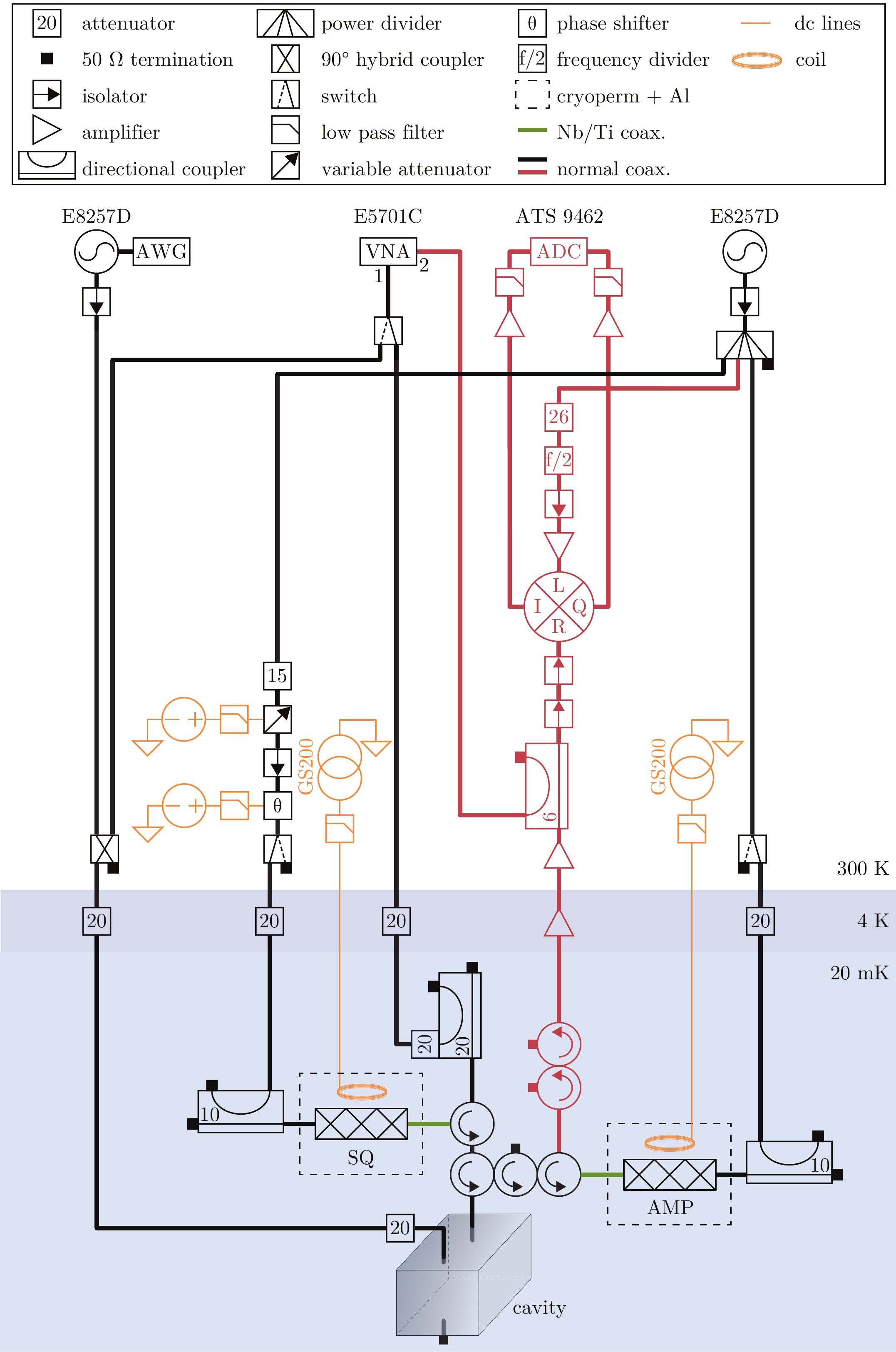}
		\caption{Full schematic of the SSR, with its room temperature control electronics. The input (black) and output (red) coaxial cables are in bold. At the $20$-mK stage, NbTi superconducting coaxial cables are shown in green, and the three adjacent circulators form a triple-junction circulator fabricated by QuinStar (QCE-070100CM30). The thinner (orange) lines represent dc cables.}
		\label{fig:fullsetup}
	\end{figure*}

	The full experimental setup is represented in Fig.\,\ref{fig:fullsetup}. The SQ-cavity-AMP ensemble is attached to the bottom plate of a dilution refrigerator. For each JPA, the pump tone is routed though a $10$-dB directional coupler connected to the pump port. The placement and configuration of these directional couplers ensures that the large pump tone required for flux pumping primarily heats up a $50$-$\ohm$ termination whose Johnson noise propagates back up the pump line, away from the JPA. A coil around each chip, connected to a dc current source at room temperature, generates a dc magnetic field, and the chip-coil ensemble is magnetically shielded with aluminum and cryoperm. Each JPA is connected to a circulator through superconducting NbTi cables in order to minimize transmission losses.
	
	An ensemble of four circulators routes the squeezed state and provides microwave isolation between the SSR elements. Two circulators between the SQ and the cavity protect the SQ from power reflected back from the cavity. These circulators provide sufficient isolation when the SQ is operated with $13$\,dB of signal gain. Similarly, three circulators separate the AMP and the cavity, as the AMP is operated with higher signal gain, around $25$\,dB. We experimentally observed that with only two circulators, undesirable feedback between the cavity and AMP perturbs the AMP's gain by effectively changing its pump's power.
	
	At the chain's input, either vacuum noise or a probe tone can be injected via a $20$-dB directional coupler. This probe tone is useful when characterizing the JPA gain profiles or when characterizing the cavity. Finally, at the output, a double-junction isolator protects the AMP from signals reflected from the next amplifier, a HEMT at $4$\,K.
	
	At room temperature, a single microwave generator (Keysight E8257D) drives both JPAs and also serves as LO for the in-phase/quadrature (IQ) mixer in the readout line. On the path leading to the SQ pump's input, a voltage-controlled variable attenuator and phase shifter provide control of both the SQ pump amplitude and phase. On the path leading to the IQ mixer, a frequency divider converts $\omega_p$ to $\omega_0$, the bare resonant frequency of both JPAs and the cavity. This divider (Pasternack PE88D2000) can input a wide range of powers (from $-20$ to $5$\,dBm) while always outputting the same power (roughly $-4$\,dBm). Thus, the AMP gain can be tuned freely with the microwave generator's output power. A second generator injects a tone through the weakly coupled port of the axion cavity. This tone can be shaped into a $9$-kHz-wide Lorentzian with an arbitrary waveform generator (AWG).
	
	At the output, a $6$\,dB directional coupler routes a fraction of the power to a vector network analyzer (VNA), which monitors signals either from the SSR chain's input or from the cavity's weakly coupled port. The VNA is used to measure \textit{in situ} the couplings $\kappa_m$ and $\kappa_l+\kappa_a$. The other portion of the output power reaches the IQ mixer's rf port. After being mixed down, the in-phase and quadrature signals are amplified and directed onto $1.9$\,MHz anti-aliasing low-pass filters, then finally digitized by an analog-to-digital converter (ADC).
	
	\subsection{Phase locking of SQ and AMP pumps}
	\label{sup:PLL}
	Enhancing the scan rate with squeezing is only beneficial if the AMP amplifies the SQ squeezed quadrature. To this end, as suggested by Fig.\,\ref{fig:squeezing}, the phase $\theta$ between SQ and AMP pumps must be maintained at $\pi/2$ (or $3\pi/2$) to a high precision, given the sharp dependence of the degree of squeezing $S$ on $\theta$. When $S>0$\,dB, the output is noisier with the SQ on than off, and the SSR is therefore detrimental to the axion search. Furthermore $S$ also depends on the SQ gain $G_s$, i.e. on the SQ pump power, and there is an optimal $G_s$ for which $S$ reaches its minimum value $S_\mathrm{min}$. In fact, as the gain increases, $S$ first improves as the squeezed state elongates in phase space, but then saturates due to distortion effects \cite{malnou2018optimal, boutin2017effect}. For this type of JPA, we experimentally find $S = S_\mathrm{min}$ at $G_s\approx13$\,dB.
	
	In order to achieve $S = S_\mathrm{min}$ throughout a data run, we implemented a feedback loop that uses the output variance $\sigma_\mathrm{on}^2$ as its control parameter on the voltage-controlled variable attenuator and phase shifter. When initializing a $9$\,hr acquisition, a 2-dimensional sweep of the variable attenuation $A_s$ and phase shift $\theta$ is used to estimate the global minimum of $\sigma_\mathrm{on}^2$. Then, periodically throughout the data run, a fast gradient descent-type algorithm corrects for small drifts of $\sigma_\mathrm{on}^2$. Note that this approach is robust to possible phase shifts due to changes in the variable attenuation and vice versa. Empirically, we are able to automatically maintain $S$ to $S_\mathrm{min}\pm0.1$\,dB, over the course of the entire experiment.
	
	We did not need to adjust the AMP pump power, as it remained stable around $25$\,dB. However, when implementing the SSR in HAYSTAC, it will have to be adjusted, in particular because the frequency of the JPAs will also be stepped in time. Note that in a practical haloscope run, sizable fluctuations of the net receiver gain in a SSR-integrated setup on timescales shorter than the raw spectrum acquisition time can be detrimental to axion detection. This is true even if all sources of added noise are overwhelmed. For our raw spectra acquisition times of $0.32$\,s, the receiver gain fluctuations are negligibly small.

	\subsection{Transmission loss between SQ and AMP}
	\label{sup:losslines}
	The presence of four circulators and several microwave connectors, including adapters from SMP (used on the JPAs chips) to SMA standards, inevitably reduces the transmission efficiency $\eta$ between SQ and AMP. We minimize $\eta$ through the use of a triple junction circulator, and superconducting SMA cables between the two JPAs and the circulators. However, $\eta$ still provides the primary limitation on the efficacy of the SSR in accelerating axionic dark matter searches.
	
	We estimate $\eta$ by measuring the output power spectral density $P_\mathrm{out}$ of vacuum fluctuations in a single quadrature amplified by the AMP with the SQ off, then repeating this measurement with the role of the JPAs interchanged \cite{malnou2018optimal}. Assuming that the gain of the amplification chain from the HEMT to the ADC stays constant, we can then deduce the microwave loss between the two JPAs as the difference in the net gain between the two cases. 
	
\begin{figure}[H]
	    \centering
    	\includegraphics[scale=0.49]{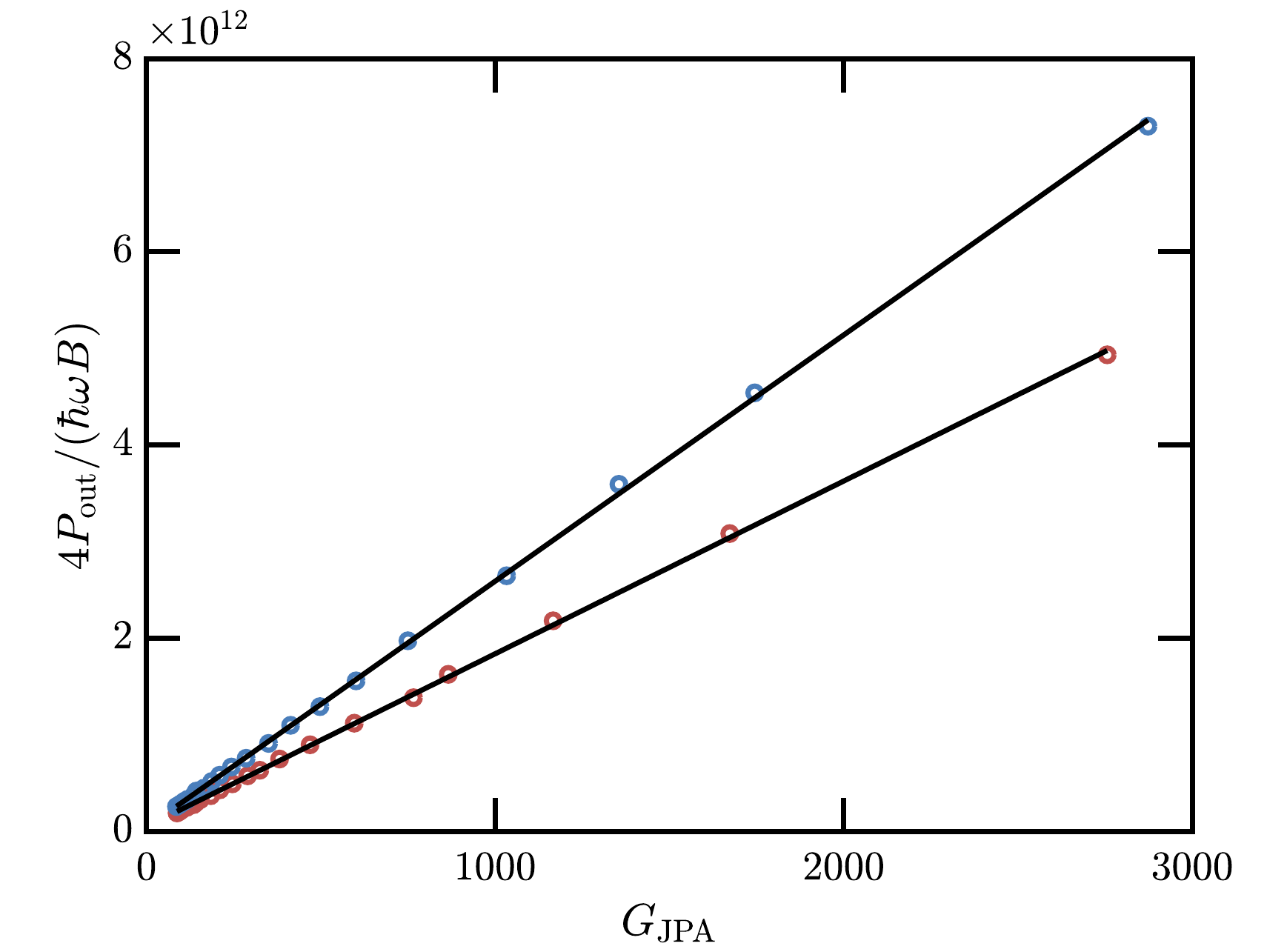}
	    \caption{Measurement of the single-quadrature output power spectral density $P_\mathrm{out}$, normalized, as a function of the JPA gain $G_\mathrm{JPA}$. Two configurations are represented: when operating only the SQ (red circles) and when operating only the AMP (blue circles). A linear fit (solid lines) to these linear responses allows us to extract $\eta$.}
	    \label{fig:outputgain}
	\end{figure}

	More precisely, having only vacuum noise at the chain's input, we have, in one quadrature:
	\begin{equation}
	    P_\mathrm{out} = \frac{1}{4}\hbar\omega B G_\mathrm{JPA}^c G_\mathrm{JPA},
	\end{equation}
	where $\omega$ is the rf angular frequency at the center of integration bandwidth $B$ for the power spectral density, $G_\mathrm{JPA}^c$ is the chain's gain after the JPA, and $G_\mathrm{JPA}$ is the SQ or AMP gain. Thus, there is a linear relation between $P_\mathrm{out}$ and $G_\mathrm{JPA}$, whose slope gives $G_\mathrm{JPA}^c$. Figure~\ref{fig:outputgain} shows $4P_\mathrm{out}/(\hbar\omega B)$, when varying either the SQ or the AMP gain. We extract $\eta = G_\mathrm{SQ}^c/G_\mathrm{AMP}^c = 0.69\pm0.01$.
	
	The efficiency $\eta$ observed in our mock-haloscope setup should not be significantly degraded by the large magnetic fields required in a real haloscope experiment such as HAYSTAC \cite{zhong2018results} or ADMX \cite{du2018search}. This is because the primary constraint imposed by the large field is an increased spatial separation between the axion cavity and the circulators and amplifiers that process the signal. This $\sim 1$-m distance is bridged with superconducting coaxial cables, whose attenuation \cite{kurpiers2017characterizing} is far subdominant to other sources of transmission loss, such as the microwave circulators in our setup, and is robust to large magnetic fields \cite{brubaker2017first}. 
	
	From $\eta$ we can estimate the squeezing $S$ that we should obtain far off cavity resonance ($\sim 10$ MHz detuned). Considering the ideal case where the reduction of the squeezed state variance is equal to the SQ single-quadrature power gain $G_s$, we obtain:
	\begin{equation}
        S = \frac{\eta}{G_s} + 1-\eta,
	\end{equation}
	which, for $G_s=13$ dB leads to $S=-4.6$ dB, in quantitative agreement with what we measure in practice.

    \section{Complementary measurements of the microwave tone's improvement in resolution}
    \label{sup:dual}
\begin{figure}[H]
    	\centering
    	\includegraphics[scale=0.49]{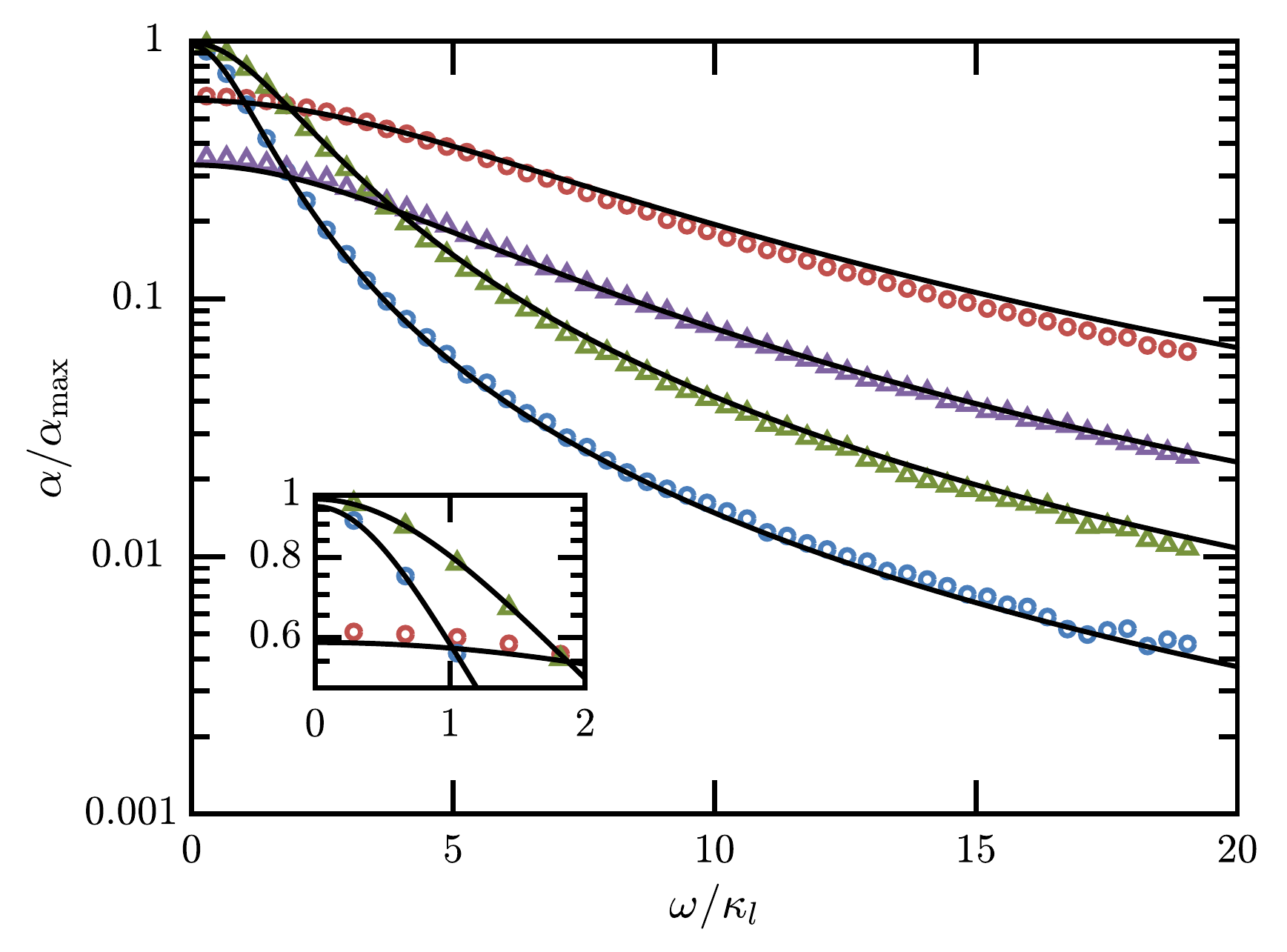}
    	\caption{Signal visibility $\alpha(\omega)$, as a function of the signal's detuning from cavity resonance. It is normalized by $\alpha_\mathrm{max}$, obtained at zero detuning for $\kappa_m=\kappa_l$. Four cases have been measured, with the corresponding theoretical predictions represented as solid lines: overcoupling ($\kappa_m=10\kappa_l$) with squeezing (red circles), near-critically coupling ($\kappa_m=1.5\kappa_l$) without squeezing (blue circles), overcoupling without squeezing (purple triangles), and near-critically coupling with squeezing (green triangles). The theory lines have been calculated for $\kappa_l=100$\,kHz, $\eta=0.69$, and $G_s=13$\,dB.}
    	\label{fig:SNRimpr_sup}
    \end{figure}

    We presented in Fig.\,\ref{fig:SNRimpr}b the visibility $\alpha(\omega)$ of a microwave tone as a function of $\omega$, the detuning between the tone's frequency and the cavity bare resonance. We compared $\alpha(\omega)$ between the two relevant cases: squeezing with a strongly overcoupled cavity versus not squeezing with a near-critically coupled cavity. Figure~\ref{fig:SNRimpr_sup} presents these, along with measurements of $\alpha(\omega)$ for the two complementary cases: strongly overcoupling without squeezing, and near-critically coupling with squeezing. In all four cases, there is excellent agreement with predictions from the theory, developed in Appendix~\ref{sup:SSRwloss} and shown in solid lines. The theory curves are not fits; we used parameter values for $\kappa_l$, $\eta$, and $G_s$ measured independently. Furthermore, the values are the same across all four cases.
    
    Squeezing is beneficial, even when not overcoupling the cavity's measurement port, as it enhances $\alpha$ off cavity resonance while having no effect on resonance. Experimentally, we obtain the complementary estimate $E^{c1}_e=1.72\pm0.03$ when squeezing and near-critically coupling, compared to the same situation without squeezing, in agreement with the theoretical prediction, $E^{c1}_t=1.77\pm0.03$. Conversely, the scan rate is worse when strongly overcoupling without squeezing, compared to near-critically coupling with squeezing: from the data, we obtain $E^{c2}_e=0.57\pm0.01$, not far from the theoretical value of $E^{c2}_t=0.52$.

	\section{Processing of the SSR spectra}
	\label{sup:processing}
	In order to detect the small cavity displacement of Sec.\,\ref{sec:faxion detection}, we acquire 401 ``raw" spectra, which we process and combine into one ``grand" spectrum. The faxion tone that we inject is sufficiently small relative to the size of the vacuum fluctuations that in any one raw spectrum it typically does not stand out. However, after processing the spectra, the faxion emerges often well above the level of vacuum noise, as in Fig.\,\ref{fig:faxionsearch}b. This appendix~provides an overview of the steps in the processing of the spectra. Our processing procedure is based closely on the work of Ref.\,\cite{brubaker2017haystac}. As such, we borrow the terminology used for the intermediate processing stages set forth in that work, and we also omit many details and rationales which are there covered extensively. 
	
	Using the setup of Fig.\,\ref{fig:schematic}, fluctuations emerging from the cavity, possibly in the presence of squeezed vacuum noise, are directed into a chain of amplifiers led by the AMP. The fluctuations are mixed down to dc, further amplified, low-pass filtered up to $1.9\ \mathrm{MHz}$, and sampled with an Alazar ATS9462 digitizer at $6\ \mathrm{MS/s}$. Each raw spectrum, distinguished as the least processed data saved to a hard disk, itself actually comprises the frequency-averaged power spectral densities of $32$ ``subspectra," each acquired over $10\ \mathrm{ms}$ and fast Fourier transformed to provide a spectral resolution of $\Delta_b = 100\ \mathrm{Hz}$. Since $401$ raw spectra are acquired for each of the $200$ squeezed and $200$ unsqueezed data runs, the live acquisition time of all the spectra  totals just over $14$ hours, not counting the dead time. In practice it took roughly $18$ hours.
	
	Once the raw spectra are in hand, they are symmetrized. Since the output of a flux-pumped JPA at a given detuning from the center of its amplification band is, in the high-gain limit, identical to the output at the opposite detuning from band center, the best one can do is to infer the spectral density measured in the homodyne configuration as coming equally from both sides of the pump. 
	
	The symmetrized spectra are then averaged according to their real frequencies (i.e. not accounting for the fictitious stepping of the cavity used to simulate a haloscope search) in order to detect excess power in either the pertinent rf or IF band \footnote{In a real haloscope search, since the local oscillator being used for homodyne measurement would be stepped along with the cavity, non-axion-induced power excesses in the rf are trickier to reject than their IF counterparts \cite{brubaker2017haystac}.}. This mean spectrum is then high-pass filtered by the profile of a Savitsky-Golay (SG) filter \cite{savitzky1964smoothing} applied to the average spectrum. A SG filter is simply a computationally quick means of applying the $d^\mathrm{th}$-degree polynomial generalization of a moving average within a window of $2W+1$ bins. If the width $W$ is much larger than the size of interesting spectral features, those features will be minimally attenuated when the spectrum is divided by the filter response, whereas features extending over wide bandwidths will be effectively removed. Following Ref.\,\cite{brubaker2017haystac}, we use $d=10$, $W=500$ for this first filtering. 
	
	The resulting real-frequency-averaged spectrum is used to detect and remove excesses of power which do not act like our faxion. Bins that exhibit power fluctuations more than $4$ standard deviations above the mean at their real frequencies are discarded along with their close neighbors as contaminated for all spectra. In practice, we keep over $99\%$ of bins, only eliminating extreme outliers. Note that since in the real frequency space, the faxion tone is being stepped, its $400$ tunings do not combine over any one narrow block of bins, and this bin rejection procedure will therefore not eliminate a faxion signal. Once contaminated bins are removed, the SG filter is recalculated, as it will be slightly different without the presence of outlier bins, and reapplied to each spectrum individually. 
	
	This baseline removal leaves a set of nearly flat ``normalized" spectra, each centered near $1$. Residual structure left from fluctuations in the overall receiver gain or its profile is then removed by dividing out the SG profile ($d = 4$, $W=500$) of each individual spectrum. The mean value of $1$ is then subtracted from each spectrum to form the ``processed" spectra, several of which are shown in Fig.\,\ref{fig:faxionsearch}a. 
	
	Next we must rescale the spectra to account for the varying sensitivity to a faxion tone as a function of its detuning from cavity resonance in any given spectrum. To form these ``rescaled" spectra, the processed spectra are divided by the relative visibility profile of the squeezed state receiver: Eq.\,\eqref{SNRloss} with an additional term accounting for the small contribution of the HEMT's added noise referred to the input of the AMP, as the weak frequency dependence of AMP gain over the bandwidth of the cavity in the presence of spectrally flat HEMT added noise contributes a correspondingly weak frequency dependence to the noise. The rescaled spectra have the important property that a given power excess from the weakly coupled port of the cavity produces a constant expectation value of rescaled spectrum power excess, regardless of detuning from the center of the cavity. As a consequence, the variance of the power distribution for each bin grows with detuning from cavity resonance. 
	
	Aligning bins along the fictitious frequencies for which the stepped faxion tone would appear stationary, it is then possible to construct the maximum likelihood estimate of the power excess in each bin. Doing so yields the ``combined" spectrum, a single spectrum whose bins contain as few as one (at the extreme edges) and as many as hundreds of contributions from individual rescaled spectra. The combined spectra bins each have a resolution of $\Delta_b$, far below the linewidth $\Delta_a$ of the faxion. The power within non-overlapping sets of $K^r = 10 \ll \Delta_a / \Delta_b$ bins are thus averaged, to form the ``rebinned" spectrum, which has $K^r \Delta_b = 1\ \mathrm{kHz}$ spectral resolution. 
	
	Finally, overlapping sets of $K^g = 41 \gg \Delta_a / K^r \Delta_b$ rebinned spectrum bins are combined, accounting for the independently experimentally determined lineshape of the faxion, in order to produce the maximum likelihood estimate of the faxion-shaped power centered on each bin. The resulting spectrum of estimated powers is the grand spectrum, shown renormalized to have mean power excess 0 and standard deviation 1 in Fig.\,\ref{fig:faxionsearch}b. In the grand spectrum, the faxion typically stands out well above the surrounding noise. 
	
	The acquisition and processing of sets of 401 raw spectra are repeated 200 times apiece for the optimal squeezed and unsqueezed cases. The two data sets require slightly different processing, chiefly because the visibility profile of the unsqueezed case does not include the frequency-dependent contribution from the squeezing. This results not from finite bandwidth effects of the SQ, but from the fact that squeezed noise is preferentially absorbed near cavity resonance by the cavity's loss port, wherefrom it is replaced with unsqueezed vacuum.
	
	Each of the 400 total acquisitions provides one measured faxion power. The powers are histogrammed for the optimal squeezed and unsqueezed cases, along with the far larger number of grand spectrum powers of bins not containing a faxion, in Fig.\,\ref{fig:faxionsearch}c. While the absolute mean values of the squeezed and unsqueezed faxion distributions carry little meaning, as they scale with our somewhat arbitrary choice of faxion power, their squared ratio, which would be preserved for a faxion of any power, gives the scan rate enhancement obtained from squeezing, as discussed in Sec.\,\ref{sec:faxion detection}. 
	
	\vspace{0.1in}
	

\begin{thebibliography}{47}%
		\makeatletter
		\providecommand \@ifxundefined [1]{%
			\@ifx{#1\undefined}
		}%
		\providecommand \@ifnum [1]{%
			\ifnum #1\expandafter \@firstoftwo
			\else \expandafter \@secondoftwo
			\fi
		}%
		\providecommand \@ifx [1]{%
			\ifx #1\expandafter \@firstoftwo
			\else \expandafter \@secondoftwo
			\fi
		}%
		\providecommand \natexlab [1]{#1}%
		\providecommand \enquote  [1]{``#1''}%
		\providecommand \bibnamefont  [1]{#1}%
		\providecommand \bibfnamefont [1]{#1}%
		\providecommand \citenamefont [1]{#1}%
		\providecommand \href@noop [0]{\@secondoftwo}%
		\providecommand \href [0]{\begingroup \@sanitize@url \@href}%
		\providecommand \@href[1]{\@@startlink{#1}\@@href}%
		\providecommand \@@href[1]{\endgroup#1\@@endlink}%
		\providecommand \@sanitize@url [0]{\catcode `\\12\catcode `\$12\catcode
			`\&12\catcode `\#12\catcode `\^12\catcode `\_12\catcode `\%12\relax}%
		\providecommand \@@startlink[1]{}%
		\providecommand \@@endlink[0]{}%
		\providecommand \url  [0]{\begingroup\@sanitize@url \@url }%
		\providecommand \@url [1]{\endgroup\@href {#1}{\urlprefix }}%
		\providecommand \urlprefix  [0]{URL }%
		\providecommand \Eprint [0]{\href }%
		\providecommand \doibase [0]{http://dx.doi.org/}%
		\providecommand \selectlanguage [0]{\@gobble}%
		\providecommand \bibinfo  [0]{\@secondoftwo}%
		\providecommand \bibfield  [0]{\@secondoftwo}%
		\providecommand \translation [1]{[#1]}%
		\providecommand \BibitemOpen [0]{}%
		\providecommand \bibitemStop [0]{}%
		\providecommand \bibitemNoStop [0]{.\EOS\space}%
		\providecommand \EOS [0]{\spacefactor3000\relax}%
		\providecommand \BibitemShut  [1]{\csname bibitem#1\endcsname}%
		\let\auto@bib@innerbib\@empty
		\bibitem [{\citenamefont {Peccei}\ and\ \citenamefont
			{Quinn}(1977{\natexlab{a}})}]{peccei1977cp}%
		\BibitemOpen
		\bibfield  {author} {\bibinfo {author} {\bibfnamefont {R.~D.}\ \bibnamefont
				{Peccei}}\ and\ \bibinfo {author} {\bibfnamefont {H.~R.}\ \bibnamefont
				{Quinn}},\ }\bibfield  {title} {\enquote {\bibinfo {title} {$\mathrm{CP}$
					conservation in the presence of pseudoparticles},}\ }\href {\doibase
			10.1103/PhysRevLett.38.1440} {\bibfield  {journal} {\bibinfo  {journal}
				{Phys. Rev. Lett.}\ }\textbf {\bibinfo {volume} {38}},\ \bibinfo {pages}
			{1440--1443} (\bibinfo {year} {1977}{\natexlab{a}})}\BibitemShut {NoStop}%
		\bibitem [{\citenamefont {Peccei}\ and\ \citenamefont
			{Quinn}(1977{\natexlab{b}})}]{peccei1977constraints}%
		\BibitemOpen
		\bibfield  {author} {\bibinfo {author} {\bibfnamefont {R.~D.}\ \bibnamefont
				{Peccei}}\ and\ \bibinfo {author} {\bibfnamefont {Helen~R.}\ \bibnamefont
				{Quinn}},\ }\bibfield  {title} {\enquote {\bibinfo {title} {Constraints
					imposed by $\mathrm{CP}$ conservation in the presence of pseudoparticles},}\
		}\href {\doibase 10.1103/PhysRevD.16.1791} {\bibfield  {journal} {\bibinfo
				{journal} {Phys. Rev. D}\ }\textbf {\bibinfo {volume} {16}},\ \bibinfo
			{pages} {1791--1797} (\bibinfo {year} {1977}{\natexlab{b}})}\BibitemShut
		{NoStop}%
		\bibitem [{\citenamefont {Preskill}\ \emph {et~al.}(1983)\citenamefont
			{Preskill}, \citenamefont {Wise},\ and\ \citenamefont
			{Wilczek}}]{preskill1983cosmology}%
		\BibitemOpen
		\bibfield  {author} {\bibinfo {author} {\bibfnamefont {John}\ \bibnamefont
				{Preskill}}, \bibinfo {author} {\bibfnamefont {Mark~B.}\ \bibnamefont
				{Wise}}, \ and\ \bibinfo {author} {\bibfnamefont {Frank}\ \bibnamefont
				{Wilczek}},\ }\bibfield  {title} {\enquote {\bibinfo {title} {Cosmology of
					the invisible axion},}\ }\href {\doibase 10.1016/0370-2693(83)90637-8}
		{\bibfield  {journal} {\bibinfo  {journal} {Phys. Lett. B}\ }\textbf
			{\bibinfo {volume} {120}},\ \bibinfo {pages} {127--132} (\bibinfo {year}
			{1983})}\BibitemShut {NoStop}%
		\bibitem [{\citenamefont {Abbott}\ and\ \citenamefont
			{Sikivie}(1983)}]{abbott1983a}%
		\BibitemOpen
		\bibfield  {author} {\bibinfo {author} {\bibfnamefont {L.F.}\ \bibnamefont
				{Abbott}}\ and\ \bibinfo {author} {\bibfnamefont {P.}~\bibnamefont
				{Sikivie}},\ }\bibfield  {title} {\enquote {\bibinfo {title} {A cosmological
					bound on the invisible axion},}\ }\href {\doibase
			https://doi.org/10.1016/0370-2693(83)90638-X} {\bibfield  {journal} {\bibinfo
				{journal} {Phys. Lett. B}\ }\textbf {\bibinfo {volume} {120}},\ \bibinfo
			{pages} {133--136} (\bibinfo {year} {1983})}\BibitemShut {NoStop}%
		\bibitem [{\citenamefont {Dine}\ and\ \citenamefont
			{Fischler}(1983)}]{dine1983the}%
		\BibitemOpen
		\bibfield  {author} {\bibinfo {author} {\bibfnamefont {Michael}\ \bibnamefont
				{Dine}}\ and\ \bibinfo {author} {\bibfnamefont {Willy}\ \bibnamefont
				{Fischler}},\ }\bibfield  {title} {\enquote {\bibinfo {title} {The
					not-so-harmless axion},}\ }\href {\doibase
			https://doi.org/10.1016/0370-2693(83)90639-1} {\bibfield  {journal} {\bibinfo
				{journal} {Phys. Lett. B}\ }\textbf {\bibinfo {volume} {120}},\ \bibinfo
			{pages} {137--141} (\bibinfo {year} {1983})}\BibitemShut {NoStop}%
		\bibitem [{\citenamefont {Caves}(1982)}]{caves1982quantum}%
		\BibitemOpen
		\bibfield  {author} {\bibinfo {author} {\bibfnamefont {Carlton~M.}\
				\bibnamefont {Caves}},\ }\bibfield  {title} {\enquote {\bibinfo {title}
				{Quantum limits on noise in linear amplifiers},}\ }\href {\doibase
			10.1103/PhysRevD.26.1817} {\bibfield  {journal} {\bibinfo  {journal} {Phys.
					Rev. D}\ }\textbf {\bibinfo {volume} {26}},\ \bibinfo {pages} {1817--1839}
			(\bibinfo {year} {1982})}\BibitemShut {NoStop}%
		\bibitem [{\citenamefont {Bradley}\ \emph {et~al.}(2003)\citenamefont
			{Bradley}, \citenamefont {Clarke}, \citenamefont {Kinion}, \citenamefont
			{Rosenberg}, \citenamefont {van Bibber}, \citenamefont {Matsuki},
			\citenamefont {M\"uck},\ and\ \citenamefont
			{Sikivie}}]{bradley2003microwave}%
		\BibitemOpen
		\bibfield  {author} {\bibinfo {author} {\bibfnamefont {Richard}\ \bibnamefont
				{Bradley}}, \bibinfo {author} {\bibfnamefont {John}\ \bibnamefont {Clarke}},
			\bibinfo {author} {\bibfnamefont {Darin}\ \bibnamefont {Kinion}}, \bibinfo
			{author} {\bibfnamefont {Leslie~J}\ \bibnamefont {Rosenberg}}, \bibinfo
			{author} {\bibfnamefont {Karl}\ \bibnamefont {van Bibber}}, \bibinfo {author}
			{\bibfnamefont {Seishi}\ \bibnamefont {Matsuki}}, \bibinfo {author}
			{\bibfnamefont {Michael}\ \bibnamefont {M\"uck}}, \ and\ \bibinfo {author}
			{\bibfnamefont {Pierre}\ \bibnamefont {Sikivie}},\ }\bibfield  {title}
		{\enquote {\bibinfo {title} {Microwave cavity searches for dark-matter
					axions},}\ }\href {\doibase 10.1103/RevModPhys.75.777} {\bibfield  {journal}
			{\bibinfo  {journal} {Rev. Mod. Phys.}\ }\textbf {\bibinfo {volume} {75}},\
			\bibinfo {pages} {777--817} (\bibinfo {year} {2003})}\BibitemShut {NoStop}%
		\bibitem [{\citenamefont {Zhitnitsky}(1980)}]{zhitnitsky1980possible}%
		\BibitemOpen
		\bibfield  {author} {\bibinfo {author} {\bibfnamefont {A.~R.}\ \bibnamefont
				{Zhitnitsky}},\ }\bibfield  {title} {\enquote {\bibinfo {title} {On possible
					suppression of the axion hadron interactions},}\ }\href@noop {} {\bibfield
			{journal} {\bibinfo  {journal} {Sov. J. Nucl. Phys.}\ }\textbf {\bibinfo
				{volume} {31}},\ \bibinfo {pages} {260} (\bibinfo {year} {1980})}\BibitemShut
		{NoStop}%
		\bibitem [{\citenamefont {Dine}\ \emph {et~al.}(1981)\citenamefont {Dine},
			\citenamefont {Fischler},\ and\ \citenamefont {Srednicki}}]{dine1981simple}%
		\BibitemOpen
		\bibfield  {author} {\bibinfo {author} {\bibfnamefont {Michael}\ \bibnamefont
				{Dine}}, \bibinfo {author} {\bibfnamefont {Willy}\ \bibnamefont {Fischler}},
			\ and\ \bibinfo {author} {\bibfnamefont {Mark}\ \bibnamefont {Srednicki}},\
		}\bibfield  {title} {\enquote {\bibinfo {title} {A simple solution to the
					strong \text{CP} problem with a harmless axion},}\ }\href {\doibase
			https://doi.org/10.1016/0370-2693(81)90590-6} {\bibfield  {journal} {\bibinfo
				{journal} {Phys. Lett. B}\ }\textbf {\bibinfo {volume} {104}},\ \bibinfo
			{pages} {199} (\bibinfo {year} {1981})}\BibitemShut {NoStop}%
		\bibitem [{\citenamefont {Du}\ \emph {et~al.}(2018)\citenamefont {Du},
			\citenamefont {Force}, \citenamefont {Khatiwada}, \citenamefont {Lentz},
			\citenamefont {Ottens}, \citenamefont {Rosenberg}, \citenamefont {Rybka},
			\citenamefont {Carosi}, \citenamefont {Woollett}, \citenamefont {Bowring},
			\citenamefont {Chou}, \citenamefont {Sonnenschein}, \citenamefont {Wester},
			\citenamefont {Boutan}, \citenamefont {Oblath}, \citenamefont {Bradley},
			\citenamefont {Daw}, \citenamefont {Dixit}, \citenamefont {Clarke},
			\citenamefont {O'Kelley}, \citenamefont {Crisosto}, \citenamefont {Gleason},
			\citenamefont {Jois}, \citenamefont {Sikivie}, \citenamefont {Stern},
			\citenamefont {Sullivan}, \citenamefont {Tanner},\ and\ \citenamefont
			{Hilton}}]{du2018search}%
		\BibitemOpen
		\bibfield  {author} {\bibinfo {author} {\bibfnamefont {N.}~\bibnamefont
				{Du}}, \bibinfo {author} {\bibfnamefont {N.}~\bibnamefont {Force}}, \bibinfo
			{author} {\bibfnamefont {R.}~\bibnamefont {Khatiwada}}, \bibinfo {author}
			{\bibfnamefont {E.}~\bibnamefont {Lentz}}, \bibinfo {author} {\bibfnamefont
				{R.}~\bibnamefont {Ottens}}, \bibinfo {author} {\bibfnamefont {L.~J}\
				\bibnamefont {Rosenberg}}, \bibinfo {author} {\bibfnamefont {G.}~\bibnamefont
				{Rybka}}, \bibinfo {author} {\bibfnamefont {G.}~\bibnamefont {Carosi}},
			\bibinfo {author} {\bibfnamefont {N.}~\bibnamefont {Woollett}}, \bibinfo
			{author} {\bibfnamefont {D.}~\bibnamefont {Bowring}}, \bibinfo {author}
			{\bibfnamefont {A.~S.}\ \bibnamefont {Chou}}, \bibinfo {author}
			{\bibfnamefont {A.}~\bibnamefont {Sonnenschein}}, \bibinfo {author}
			{\bibfnamefont {W.}~\bibnamefont {Wester}}, \bibinfo {author} {\bibfnamefont
				{C.}~\bibnamefont {Boutan}}, \bibinfo {author} {\bibfnamefont {N.~S.}\
				\bibnamefont {Oblath}}, \bibinfo {author} {\bibfnamefont {R.}~\bibnamefont
				{Bradley}}, \bibinfo {author} {\bibfnamefont {E.~J.}\ \bibnamefont {Daw}},
			\bibinfo {author} {\bibfnamefont {A.~V.}\ \bibnamefont {Dixit}}, \bibinfo
			{author} {\bibfnamefont {J.}~\bibnamefont {Clarke}}, \bibinfo {author}
			{\bibfnamefont {S.~R.}\ \bibnamefont {O'Kelley}}, \bibinfo {author}
			{\bibfnamefont {N.}~\bibnamefont {Crisosto}}, \bibinfo {author}
			{\bibfnamefont {J.~R.}\ \bibnamefont {Gleason}}, \bibinfo {author}
			{\bibfnamefont {S.}~\bibnamefont {Jois}}, \bibinfo {author} {\bibfnamefont
				{P.}~\bibnamefont {Sikivie}}, \bibinfo {author} {\bibfnamefont
				{I.}~\bibnamefont {Stern}}, \bibinfo {author} {\bibfnamefont {N.~S.}\
				\bibnamefont {Sullivan}}, \bibinfo {author} {\bibfnamefont {D.~B}\
				\bibnamefont {Tanner}}, \ and\ \bibinfo {author} {\bibfnamefont {G.~C.}\
				\bibnamefont {Hilton}} (\bibinfo {collaboration} {ADMX Collaboration}),\
		}\bibfield  {title} {\enquote {\bibinfo {title} {Search for invisible axion
					dark matter with the axion dark matter experiment},}\ }\href {\doibase
			10.1103/PhysRevLett.120.151301} {\bibfield  {journal} {\bibinfo  {journal}
				{Phys. Rev. Lett.}\ }\textbf {\bibinfo {volume} {120}},\ \bibinfo {pages}
			{151301} (\bibinfo {year} {2018})}\BibitemShut {NoStop}%
		\bibitem [{\citenamefont {Zhong}\ \emph {et~al.}(2018)\citenamefont {Zhong},
			\citenamefont {Al~Kenany}, \citenamefont {Backes}, \citenamefont {Brubaker},
			\citenamefont {Cahn}, \citenamefont {Carosi}, \citenamefont {Gurevich},
			\citenamefont {Kindel}, \citenamefont {Lamoreaux}, \citenamefont {Lehnert},
			\citenamefont {Lewis}, \citenamefont {Malnou}, \citenamefont {Maruyama},
			\citenamefont {Palken}, \citenamefont {Rapidis}, \citenamefont {Root},
			\citenamefont {Simanovskaia}, \citenamefont {Shokair}, \citenamefont
			{Speller}, \citenamefont {Urdinaran},\ and\ \citenamefont {van
				Bibber}}]{zhong2018results}%
		\BibitemOpen
		\bibfield  {author} {\bibinfo {author} {\bibfnamefont {L.}~\bibnamefont
				{Zhong}}, \bibinfo {author} {\bibfnamefont {S.}~\bibnamefont {Al~Kenany}},
			\bibinfo {author} {\bibfnamefont {K.~M.}\ \bibnamefont {Backes}}, \bibinfo
			{author} {\bibfnamefont {B.~M.}\ \bibnamefont {Brubaker}}, \bibinfo {author}
			{\bibfnamefont {S.~B.}\ \bibnamefont {Cahn}}, \bibinfo {author}
			{\bibfnamefont {G.}~\bibnamefont {Carosi}}, \bibinfo {author} {\bibfnamefont
				{Y.~V.}\ \bibnamefont {Gurevich}}, \bibinfo {author} {\bibfnamefont {W.~F.}\
				\bibnamefont {Kindel}}, \bibinfo {author} {\bibfnamefont {S.~K.}\
				\bibnamefont {Lamoreaux}}, \bibinfo {author} {\bibfnamefont {K.~W.}\
				\bibnamefont {Lehnert}}, \bibinfo {author} {\bibfnamefont {S.~M.}\
				\bibnamefont {Lewis}}, \bibinfo {author} {\bibfnamefont {M.}~\bibnamefont
				{Malnou}}, \bibinfo {author} {\bibfnamefont {R.~H.}\ \bibnamefont
				{Maruyama}}, \bibinfo {author} {\bibfnamefont {D.~A.}\ \bibnamefont
				{Palken}}, \bibinfo {author} {\bibfnamefont {N.~M.}\ \bibnamefont {Rapidis}},
			\bibinfo {author} {\bibfnamefont {J.~R.}\ \bibnamefont {Root}}, \bibinfo
			{author} {\bibfnamefont {M.}~\bibnamefont {Simanovskaia}}, \bibinfo {author}
			{\bibfnamefont {T.~M.}\ \bibnamefont {Shokair}}, \bibinfo {author}
			{\bibfnamefont {D.~H.}\ \bibnamefont {Speller}}, \bibinfo {author}
			{\bibfnamefont {I.}~\bibnamefont {Urdinaran}}, \ and\ \bibinfo {author}
			{\bibfnamefont {K.~A.}\ \bibnamefont {van Bibber}},\ }\bibfield  {title}
		{\enquote {\bibinfo {title} {Results from phase 1 of the {HAYSTAC} microwave
					cavity axion experiment},}\ }\href {\doibase 10.1103/PhysRevD.97.092001}
		{\bibfield  {journal} {\bibinfo  {journal} {Phys. Rev. D}\ }\textbf {\bibinfo
				{volume} {97}},\ \bibinfo {pages} {092001} (\bibinfo {year}
			{2018})}\BibitemShut {NoStop}%
		\bibitem [{\citenamefont {Caves}(1981)}]{caves1981quantum}%
		\BibitemOpen
		\bibfield  {author} {\bibinfo {author} {\bibfnamefont {Carlton~M.}\
				\bibnamefont {Caves}},\ }\bibfield  {title} {\enquote {\bibinfo {title}
				{Quantum-mechanical noise in an interferometer},}\ }\href {\doibase
			10.1103/PhysRevD.23.1693} {\bibfield  {journal} {\bibinfo  {journal} {Phys.
					Rev. D}\ }\textbf {\bibinfo {volume} {23}},\ \bibinfo {pages} {1693--1708}
			(\bibinfo {year} {1981})}\BibitemShut {NoStop}%
		\bibitem [{\citenamefont {Collaboration}(2011)}]{abadie2011gravitational}%
		\BibitemOpen
		\bibfield  {author} {\bibinfo {author} {\bibfnamefont {The LIGO~Scientific}\
				\bibnamefont {Collaboration}},\ }\bibfield  {title} {\enquote {\bibinfo
				{title} {A gravitational wave observatory operating beyond the quantum
					shot-noise limit},}\ }\href {\doibase 10.1038/nphys2083} {\bibfield
			{journal} {\bibinfo  {journal} {Nat. Phys.}\ }\textbf {\bibinfo {volume}
				{7}},\ \bibinfo {pages} {962} (\bibinfo {year} {2011})}\BibitemShut {NoStop}%
		\bibitem [{\citenamefont {Aasi}\ \emph {et~al.}(2013)\citenamefont {Aasi},
			\citenamefont {Abadie}, \citenamefont {Abbott}, \citenamefont {Abbott},
			\citenamefont {Abbott}, \citenamefont {Abernathy}, \citenamefont {Adams},
			\citenamefont {Adams}, \citenamefont {Addesso}, \citenamefont {Adhikari}
			\emph {et~al.}}]{aasi2013enhanced}%
		\BibitemOpen
		\bibfield  {author} {\bibinfo {author} {\bibfnamefont {J.}~\bibnamefont
				{Aasi}}, \bibinfo {author} {\bibfnamefont {J.}~\bibnamefont {Abadie}},
			\bibinfo {author} {\bibfnamefont {B.~P.}\ \bibnamefont {Abbott}}, \bibinfo
			{author} {\bibfnamefont {R.}~\bibnamefont {Abbott}}, \bibinfo {author}
			{\bibfnamefont {T.~D.}\ \bibnamefont {Abbott}}, \bibinfo {author}
			{\bibfnamefont {M.~R.}\ \bibnamefont {Abernathy}}, \bibinfo {author}
			{\bibfnamefont {C.}~\bibnamefont {Adams}}, \bibinfo {author} {\bibfnamefont
				{T.}~\bibnamefont {Adams}}, \bibinfo {author} {\bibfnamefont
				{P.}~\bibnamefont {Addesso}}, \bibinfo {author} {\bibfnamefont {R.~X.}\
				\bibnamefont {Adhikari}},  \emph {et~al.},\ }\bibfield  {title} {\enquote
			{\bibinfo {title} {Enhanced sensitivity of the {LIGO} gravitational wave
					detector by using squeezed states of light},}\ }\href {\doibase
			10.1038/nphoton.2013.177} {\bibfield  {journal} {\bibinfo  {journal} {Nat.
					Photon.}\ }\textbf {\bibinfo {volume} {7}},\ \bibinfo {pages} {613} (\bibinfo
			{year} {2013})}\BibitemShut {NoStop}%
		\bibitem [{\citenamefont {Kimble}\ \emph {et~al.}(2001)\citenamefont {Kimble},
			\citenamefont {Levin}, \citenamefont {Matsko}, \citenamefont {Thorne},\ and\
			\citenamefont {Vyatchanin}}]{kimble2001conversion}%
		\BibitemOpen
		\bibfield  {author} {\bibinfo {author} {\bibfnamefont {H.~J.}\ \bibnamefont
				{Kimble}}, \bibinfo {author} {\bibfnamefont {Yuri}\ \bibnamefont {Levin}},
			\bibinfo {author} {\bibfnamefont {Andrey~B.}\ \bibnamefont {Matsko}},
			\bibinfo {author} {\bibfnamefont {Kip~S.}\ \bibnamefont {Thorne}}, \ and\
			\bibinfo {author} {\bibfnamefont {Sergey~P.}\ \bibnamefont {Vyatchanin}},\
		}\bibfield  {title} {\enquote {\bibinfo {title} {Conversion of conventional
					gravitational-wave interferometers into quantum nondemolition interferometers
					by modifying their input and/or output optics},}\ }\href {\doibase
			10.1103/PhysRevD.65.022002} {\bibfield  {journal} {\bibinfo  {journal} {Phys.
					Rev. D}\ }\textbf {\bibinfo {volume} {65}},\ \bibinfo {pages} {022002}
			(\bibinfo {year} {2001})}\BibitemShut {NoStop}%
		\bibitem [{\citenamefont {Murch}\ \emph {et~al.}(2013)\citenamefont {Murch},
			\citenamefont {Weber}, \citenamefont {Beck}, \citenamefont {Ginossar},\ and\
			\citenamefont {Siddiqi}}]{murch2013reduction}%
		\BibitemOpen
		\bibfield  {author} {\bibinfo {author} {\bibfnamefont {K.~W.}\ \bibnamefont
				{Murch}}, \bibinfo {author} {\bibfnamefont {S.~J.}\ \bibnamefont {Weber}},
			\bibinfo {author} {\bibfnamefont {K.~M.}\ \bibnamefont {Beck}}, \bibinfo
			{author} {\bibfnamefont {E.}~\bibnamefont {Ginossar}}, \ and\ \bibinfo
			{author} {\bibfnamefont {I.}~\bibnamefont {Siddiqi}},\ }\bibfield  {title}
		{\enquote {\bibinfo {title} {Reduction of the radiative decay of atomic
					coherence in squeezed vacuum},}\ }\href
		{http://www.nature.com/nature/journal/v499/n7456/full/nature12264.html
			http://arxiv.org/abs/1301.6276} {\bibfield  {journal} {\bibinfo  {journal}
				{Nature}\ }\textbf {\bibinfo {volume} {499}},\ \bibinfo {pages} {62--65}
			(\bibinfo {year} {2013})}\BibitemShut {NoStop}%
		\bibitem [{\citenamefont {Clark}\ \emph {et~al.}(2016)\citenamefont {Clark},
			\citenamefont {Lecocq}, \citenamefont {Simmonds}, \citenamefont {Aumentado},\
			and\ \citenamefont {Teufel}}]{clark2016observation}%
		\BibitemOpen
		\bibfield  {author} {\bibinfo {author} {\bibfnamefont {Jeremy~B}\
				\bibnamefont {Clark}}, \bibinfo {author} {\bibfnamefont {Florent}\
				\bibnamefont {Lecocq}}, \bibinfo {author} {\bibfnamefont {Raymond~W}\
				\bibnamefont {Simmonds}}, \bibinfo {author} {\bibfnamefont {Jos{\'e}}\
				\bibnamefont {Aumentado}}, \ and\ \bibinfo {author} {\bibfnamefont {John~D}\
				\bibnamefont {Teufel}},\ }\bibfield  {title} {\enquote {\bibinfo {title}
				{Observation of strong radiation pressure forces from squeezed light on a
					mechanical oscillator},}\ }\href {http://dx.doi.org/10.1038/nphys3701}
		{\bibfield  {journal} {\bibinfo  {journal} {Nat. Phys.}\ }\textbf {\bibinfo
				{volume} {12}},\ \bibinfo {pages} {683--687} (\bibinfo {year}
			{2016})}\BibitemShut {NoStop}%
		\bibitem [{\citenamefont {Bienfait}\ \emph {et~al.}(2017)\citenamefont
			{Bienfait}, \citenamefont {Campagne-Ibarcq}, \citenamefont {Kiilerich},
			\citenamefont {Zhou}, \citenamefont {Probst}, \citenamefont {Pla},
			\citenamefont {Schenkel}, \citenamefont {Vion}, \citenamefont {Esteve},
			\citenamefont {Morton}, \citenamefont {Moelmer},\ and\ \citenamefont
			{Bertet}}]{bienfait2017magnetic}%
		\BibitemOpen
		\bibfield  {author} {\bibinfo {author} {\bibfnamefont {A.}~\bibnamefont
				{Bienfait}}, \bibinfo {author} {\bibfnamefont {P.}~\bibnamefont
				{Campagne-Ibarcq}}, \bibinfo {author} {\bibfnamefont {A.~H.}\ \bibnamefont
				{Kiilerich}}, \bibinfo {author} {\bibfnamefont {X.}~\bibnamefont {Zhou}},
			\bibinfo {author} {\bibfnamefont {S.}~\bibnamefont {Probst}}, \bibinfo
			{author} {\bibfnamefont {J.~J.}\ \bibnamefont {Pla}}, \bibinfo {author}
			{\bibfnamefont {T.}~\bibnamefont {Schenkel}}, \bibinfo {author}
			{\bibfnamefont {D.}~\bibnamefont {Vion}}, \bibinfo {author} {\bibfnamefont
				{D.}~\bibnamefont {Esteve}}, \bibinfo {author} {\bibfnamefont {J.~J.~L.}\
				\bibnamefont {Morton}}, \bibinfo {author} {\bibfnamefont {K.}~\bibnamefont
				{Moelmer}}, \ and\ \bibinfo {author} {\bibfnamefont {P.}~\bibnamefont
				{Bertet}},\ }\bibfield  {title} {\enquote {\bibinfo {title} {Magnetic
					resonance with squeezed microwaves},}\ }\href {\doibase
			10.1103/PhysRevX.7.041011} {\bibfield  {journal} {\bibinfo  {journal} {Phys.
					Rev. X}\ }\textbf {\bibinfo {volume} {7}},\ \bibinfo {pages} {041011}
			(\bibinfo {year} {2017})}\BibitemShut {NoStop}%
		\bibitem [{\citenamefont {Sikivie}(1983)}]{sikivie1983experimental}%
		\BibitemOpen
		\bibfield  {author} {\bibinfo {author} {\bibfnamefont {P.}~\bibnamefont
				{Sikivie}},\ }\bibfield  {title} {\enquote {\bibinfo {title} {Experimental
					tests of the "invisible" axion},}\ }\href {\doibase
			10.1103/PhysRevLett.51.1415} {\bibfield  {journal} {\bibinfo  {journal}
				{Phys. Rev. Lett.}\ }\textbf {\bibinfo {volume} {51}},\ \bibinfo {pages}
			{1415--1417} (\bibinfo {year} {1983})}\BibitemShut {NoStop}%
		\bibitem [{\citenamefont {Yamamoto}\ \emph {et~al.}(2008)\citenamefont
			{Yamamoto}, \citenamefont {Inomata}, \citenamefont {Watanabe}, \citenamefont
			{Matsuba}, \citenamefont {Miyazaki}, \citenamefont {Oliver}, \citenamefont
			{Nakamura},\ and\ \citenamefont {Tsai}}]{yamamoto2008flux}%
		\BibitemOpen
		\bibfield  {author} {\bibinfo {author} {\bibfnamefont {T.}~\bibnamefont
				{Yamamoto}}, \bibinfo {author} {\bibfnamefont {K.}~\bibnamefont {Inomata}},
			\bibinfo {author} {\bibfnamefont {M.}~\bibnamefont {Watanabe}}, \bibinfo
			{author} {\bibfnamefont {K.}~\bibnamefont {Matsuba}}, \bibinfo {author}
			{\bibfnamefont {T.}~\bibnamefont {Miyazaki}}, \bibinfo {author}
			{\bibfnamefont {W.~D.}\ \bibnamefont {Oliver}}, \bibinfo {author}
			{\bibfnamefont {Y.}~\bibnamefont {Nakamura}}, \ and\ \bibinfo {author}
			{\bibfnamefont {J.~S.}\ \bibnamefont {Tsai}},\ }\bibfield  {title} {\enquote
			{\bibinfo {title} {Flux-driven {J}osephson parametric amplifier},}\ }\href
		{\doibase 10.1063/1.2964182} {\bibfield  {journal} {\bibinfo  {journal}
				{Appl. Phys. Lett.}\ }\textbf {\bibinfo {volume} {93}},\ \bibinfo {pages}
			{042510} (\bibinfo {year} {2008})}\BibitemShut {NoStop}%
		\bibitem [{\citenamefont {Castellanos-Beltran}\ \emph
			{et~al.}(2008)\citenamefont {Castellanos-Beltran}, \citenamefont {Irwin},
			\citenamefont {Hilton}, \citenamefont {Vale},\ and\ \citenamefont
			{Lehnert}}]{castellanos2008amplification}%
		\BibitemOpen
		\bibfield  {author} {\bibinfo {author} {\bibfnamefont {M.~A.}\ \bibnamefont
				{Castellanos-Beltran}}, \bibinfo {author} {\bibfnamefont {K.~D.}\
				\bibnamefont {Irwin}}, \bibinfo {author} {\bibfnamefont {G.~C.}\ \bibnamefont
				{Hilton}}, \bibinfo {author} {\bibfnamefont {L.~R.}\ \bibnamefont {Vale}}, \
			and\ \bibinfo {author} {\bibfnamefont {K.~W.}\ \bibnamefont {Lehnert}},\
		}\bibfield  {title} {\enquote {\bibinfo {title} {Amplification and squeezing
					of quantum noise with a tunable {J}osephson metamaterial},}\ }\href
		{http://dx.doi.org/10.1038/nphys1090} {\bibfield  {journal} {\bibinfo
				{journal} {Nat. Phys.}\ }\textbf {\bibinfo {volume} {4}},\ \bibinfo {pages}
			{929--931} (\bibinfo {year} {2008})}\BibitemShut {NoStop}%
		\bibitem [{\citenamefont {Zhou}\ \emph {et~al.}(2014)\citenamefont {Zhou},
			\citenamefont {Schmitt}, \citenamefont {Bertet}, \citenamefont {Vion},
			\citenamefont {Wustmann}, \citenamefont {Shumeiko},\ and\ \citenamefont
			{Esteve}}]{zhou2014high}%
		\BibitemOpen
		\bibfield  {author} {\bibinfo {author} {\bibfnamefont {X.}~\bibnamefont
				{Zhou}}, \bibinfo {author} {\bibfnamefont {V.}~\bibnamefont {Schmitt}},
			\bibinfo {author} {\bibfnamefont {P.}~\bibnamefont {Bertet}}, \bibinfo
			{author} {\bibfnamefont {D.}~\bibnamefont {Vion}}, \bibinfo {author}
			{\bibfnamefont {W.}~\bibnamefont {Wustmann}}, \bibinfo {author}
			{\bibfnamefont {V.}~\bibnamefont {Shumeiko}}, \ and\ \bibinfo {author}
			{\bibfnamefont {D.}~\bibnamefont {Esteve}},\ }\bibfield  {title} {\enquote
			{\bibinfo {title} {High-gain weakly nonlinear flux-modulated {J}osephson
					parametric amplifier using a {SQUID} array},}\ }\href {\doibase
			10.1103/PhysRevB.89.214517} {\bibfield  {journal} {\bibinfo  {journal} {Phys.
					Rev. B}\ }\textbf {\bibinfo {volume} {89}},\ \bibinfo {pages} {214517}
			(\bibinfo {year} {2014})}\BibitemShut {NoStop}%
		\bibitem [{\citenamefont {Brubaker}\ \emph
			{et~al.}(2017{\natexlab{a}})\citenamefont {Brubaker}, \citenamefont {Zhong},
			\citenamefont {Gurevich}, \citenamefont {Cahn}, \citenamefont {Lamoreaux},
			\citenamefont {Simanovskaia}, \citenamefont {Root}, \citenamefont {Lewis},
			\citenamefont {Al~Kenany}, \citenamefont {Backes}, \citenamefont {Urdinaran},
			\citenamefont {Rapidis}, \citenamefont {Shokair}, \citenamefont {van Bibber},
			\citenamefont {Palken}, \citenamefont {Malnou}, \citenamefont {Kindel},
			\citenamefont {Anil}, \citenamefont {Lehnert},\ and\ \citenamefont
			{Carosi}}]{brubaker2017first}%
		\BibitemOpen
		\bibfield  {author} {\bibinfo {author} {\bibfnamefont {B.~M.}\ \bibnamefont
				{Brubaker}}, \bibinfo {author} {\bibfnamefont {L.}~\bibnamefont {Zhong}},
			\bibinfo {author} {\bibfnamefont {Y.~V.}\ \bibnamefont {Gurevich}}, \bibinfo
			{author} {\bibfnamefont {S.~B.}\ \bibnamefont {Cahn}}, \bibinfo {author}
			{\bibfnamefont {S.~K.}\ \bibnamefont {Lamoreaux}}, \bibinfo {author}
			{\bibfnamefont {M.}~\bibnamefont {Simanovskaia}}, \bibinfo {author}
			{\bibfnamefont {J.~R.}\ \bibnamefont {Root}}, \bibinfo {author}
			{\bibfnamefont {S.~M.}\ \bibnamefont {Lewis}}, \bibinfo {author}
			{\bibfnamefont {S.}~\bibnamefont {Al~Kenany}}, \bibinfo {author}
			{\bibfnamefont {K.~M.}\ \bibnamefont {Backes}}, \bibinfo {author}
			{\bibfnamefont {I.}~\bibnamefont {Urdinaran}}, \bibinfo {author}
			{\bibfnamefont {N.~M.}\ \bibnamefont {Rapidis}}, \bibinfo {author}
			{\bibfnamefont {T.~M.}\ \bibnamefont {Shokair}}, \bibinfo {author}
			{\bibfnamefont {K.~A.}\ \bibnamefont {van Bibber}}, \bibinfo {author}
			{\bibfnamefont {D.~A.}\ \bibnamefont {Palken}}, \bibinfo {author}
			{\bibfnamefont {M.}~\bibnamefont {Malnou}}, \bibinfo {author} {\bibfnamefont
				{W.~F.}\ \bibnamefont {Kindel}}, \bibinfo {author} {\bibfnamefont {M.~A.}\
				\bibnamefont {Anil}}, \bibinfo {author} {\bibfnamefont {K.~W.}\ \bibnamefont
				{Lehnert}}, \ and\ \bibinfo {author} {\bibfnamefont {G.}~\bibnamefont
				{Carosi}},\ }\bibfield  {title} {\enquote {\bibinfo {title} {First results
					from a microwave cavity axion search at $24\text{ }\text{
					}\ensuremath{\mu}\mathrm{eV}$},}\ }\href {\doibase
			10.1103/PhysRevLett.118.061302} {\bibfield  {journal} {\bibinfo  {journal}
				{Phys. Rev. Lett.}\ }\textbf {\bibinfo {volume} {118}},\ \bibinfo {pages}
			{061302} (\bibinfo {year} {2017}{\natexlab{a}})}\BibitemShut {NoStop}%
		\bibitem [{\citenamefont {Mallet}\ \emph {et~al.}(2011)\citenamefont {Mallet},
			\citenamefont {Castellanos-Beltran}, \citenamefont {Ku}, \citenamefont
			{Glancy}, \citenamefont {Knill}, \citenamefont {Irwin}, \citenamefont
			{Hilton}, \citenamefont {Vale},\ and\ \citenamefont
			{Lehnert}}]{mallet2011quantum}%
		\BibitemOpen
		\bibfield  {author} {\bibinfo {author} {\bibfnamefont {F.}~\bibnamefont
				{Mallet}}, \bibinfo {author} {\bibfnamefont {M.~A.}\ \bibnamefont
				{Castellanos-Beltran}}, \bibinfo {author} {\bibfnamefont {H.~S.}\
				\bibnamefont {Ku}}, \bibinfo {author} {\bibfnamefont {S.}~\bibnamefont
				{Glancy}}, \bibinfo {author} {\bibfnamefont {E.}~\bibnamefont {Knill}},
			\bibinfo {author} {\bibfnamefont {K.~D.}\ \bibnamefont {Irwin}}, \bibinfo
			{author} {\bibfnamefont {G.~C.}\ \bibnamefont {Hilton}}, \bibinfo {author}
			{\bibfnamefont {L.~R.}\ \bibnamefont {Vale}}, \ and\ \bibinfo {author}
			{\bibfnamefont {K.~W.}\ \bibnamefont {Lehnert}},\ }\bibfield  {title}
		{\enquote {\bibinfo {title} {Quantum state tomography of an itinerant
					squeezed microwave field},}\ }\href {\doibase 10.1103/PhysRevLett.106.220502}
		{\bibfield  {journal} {\bibinfo  {journal} {Phys. Rev. Lett.}\ }\textbf
			{\bibinfo {volume} {106}},\ \bibinfo {pages} {220502} (\bibinfo {year}
			{2011})}\BibitemShut {NoStop}%
		\bibitem [{\citenamefont {Menzel}\ \emph {et~al.}(2012)\citenamefont {Menzel},
			\citenamefont {Di~Candia}, \citenamefont {Deppe}, \citenamefont {Eder},
			\citenamefont {Zhong}, \citenamefont {Ihmig}, \citenamefont {Haeberlein},
			\citenamefont {Baust}, \citenamefont {Hoffmann}, \citenamefont {Ballester},
			\citenamefont {Inomata}, \citenamefont {Yamamoto}, \citenamefont {Nakamura},
			\citenamefont {Solano}, \citenamefont {Marx},\ and\ \citenamefont
			{Gross}}]{menzel2012path}%
		\BibitemOpen
		\bibfield  {author} {\bibinfo {author} {\bibfnamefont {E.~P.}\ \bibnamefont
				{Menzel}}, \bibinfo {author} {\bibfnamefont {R.}~\bibnamefont {Di~Candia}},
			\bibinfo {author} {\bibfnamefont {F.}~\bibnamefont {Deppe}}, \bibinfo
			{author} {\bibfnamefont {P.}~\bibnamefont {Eder}}, \bibinfo {author}
			{\bibfnamefont {L.}~\bibnamefont {Zhong}}, \bibinfo {author} {\bibfnamefont
				{M.}~\bibnamefont {Ihmig}}, \bibinfo {author} {\bibfnamefont
				{M.}~\bibnamefont {Haeberlein}}, \bibinfo {author} {\bibfnamefont
				{A.}~\bibnamefont {Baust}}, \bibinfo {author} {\bibfnamefont
				{E.}~\bibnamefont {Hoffmann}}, \bibinfo {author} {\bibfnamefont
				{D.}~\bibnamefont {Ballester}}, \bibinfo {author} {\bibfnamefont
				{K.}~\bibnamefont {Inomata}}, \bibinfo {author} {\bibfnamefont
				{T.}~\bibnamefont {Yamamoto}}, \bibinfo {author} {\bibfnamefont
				{Y.}~\bibnamefont {Nakamura}}, \bibinfo {author} {\bibfnamefont
				{E.}~\bibnamefont {Solano}}, \bibinfo {author} {\bibfnamefont
				{A.}~\bibnamefont {Marx}}, \ and\ \bibinfo {author} {\bibfnamefont
				{R.}~\bibnamefont {Gross}},\ }\bibfield  {title} {\enquote {\bibinfo {title}
				{Path entanglement of continuous-variable quantum microwaves},}\ }\href
		{\doibase 10.1103/PhysRevLett.109.250502} {\bibfield  {journal} {\bibinfo
				{journal} {Phys. Rev. Lett.}\ }\textbf {\bibinfo {volume} {109}},\ \bibinfo
			{pages} {250502} (\bibinfo {year} {2012})}\BibitemShut {NoStop}%
		\bibitem [{\citenamefont {Boutin}\ \emph {et~al.}(2017)\citenamefont {Boutin},
			\citenamefont {Toyli}, \citenamefont {Venkatramani}, \citenamefont {Eddins},
			\citenamefont {Siddiqi},\ and\ \citenamefont {Blais}}]{boutin2017effect}%
		\BibitemOpen
		\bibfield  {author} {\bibinfo {author} {\bibfnamefont {Samuel}\ \bibnamefont
				{Boutin}}, \bibinfo {author} {\bibfnamefont {David~M.}\ \bibnamefont
				{Toyli}}, \bibinfo {author} {\bibfnamefont {Aditya~V.}\ \bibnamefont
				{Venkatramani}}, \bibinfo {author} {\bibfnamefont {Andrew~W.}\ \bibnamefont
				{Eddins}}, \bibinfo {author} {\bibfnamefont {Irfan}\ \bibnamefont {Siddiqi}},
			\ and\ \bibinfo {author} {\bibfnamefont {Alexandre}\ \bibnamefont {Blais}},\
		}\bibfield  {title} {\enquote {\bibinfo {title} {Effect of higher-order
					nonlinearities on amplification and squeezing in {J}osephson parametric
					amplifiers},}\ }\href {\doibase 10.1103/PhysRevApplied.8.054030} {\bibfield
			{journal} {\bibinfo  {journal} {Phys. Rev. Appl.}\ }\textbf {\bibinfo
				{volume} {8}},\ \bibinfo {pages} {054030} (\bibinfo {year}
			{2017})}\BibitemShut {NoStop}%
		\bibitem [{\citenamefont {Malnou}\ \emph {et~al.}(2018)\citenamefont {Malnou},
			\citenamefont {Palken}, \citenamefont {Vale}, \citenamefont {Hilton},\ and\
			\citenamefont {Lehnert}}]{malnou2018optimal}%
		\BibitemOpen
		\bibfield  {author} {\bibinfo {author} {\bibfnamefont {M.}~\bibnamefont
				{Malnou}}, \bibinfo {author} {\bibfnamefont {D.~A.}\ \bibnamefont {Palken}},
			\bibinfo {author} {\bibfnamefont {Leila~R.}\ \bibnamefont {Vale}}, \bibinfo
			{author} {\bibfnamefont {Gene~C.}\ \bibnamefont {Hilton}}, \ and\ \bibinfo
			{author} {\bibfnamefont {K.~W.}\ \bibnamefont {Lehnert}},\ }\bibfield
		{title} {\enquote {\bibinfo {title} {Optimal operation of a {J}osephson
					parametric amplifier for vacuum squeezing},}\ }\href {\doibase
			10.1103/PhysRevApplied.9.044023} {\bibfield  {journal} {\bibinfo  {journal}
				{Phys. Rev. Appl.}\ }\textbf {\bibinfo {volume} {9}},\ \bibinfo {pages}
			{044023} (\bibinfo {year} {2018})}\BibitemShut {NoStop}%
		\bibitem [{\citenamefont {Pogorzalek}\ \emph {et~al.}(2017)\citenamefont
			{Pogorzalek}, \citenamefont {Fedorov}, \citenamefont {Zhong}, \citenamefont
			{Goetz}, \citenamefont {Wulschner}, \citenamefont {Fischer}, \citenamefont
			{Eder}, \citenamefont {Xie}, \citenamefont {Inomata}, \citenamefont
			{Yamamoto}, \citenamefont {Nakamura}, \citenamefont {Marx}, \citenamefont
			{Deppe},\ and\ \citenamefont {Gross}}]{pogorzalek2017hysteretic}%
		\BibitemOpen
		\bibfield  {author} {\bibinfo {author} {\bibfnamefont {Stefan}\ \bibnamefont
				{Pogorzalek}}, \bibinfo {author} {\bibfnamefont {Kirill~G.}\ \bibnamefont
				{Fedorov}}, \bibinfo {author} {\bibfnamefont {Ling}\ \bibnamefont {Zhong}},
			\bibinfo {author} {\bibfnamefont {Jan}\ \bibnamefont {Goetz}}, \bibinfo
			{author} {\bibfnamefont {Friedrich}\ \bibnamefont {Wulschner}}, \bibinfo
			{author} {\bibfnamefont {Michael}\ \bibnamefont {Fischer}}, \bibinfo {author}
			{\bibfnamefont {Peter}\ \bibnamefont {Eder}}, \bibinfo {author}
			{\bibfnamefont {Edwar}\ \bibnamefont {Xie}}, \bibinfo {author} {\bibfnamefont
				{Kunihiro}\ \bibnamefont {Inomata}}, \bibinfo {author} {\bibfnamefont
				{Tsuyoshi}\ \bibnamefont {Yamamoto}}, \bibinfo {author} {\bibfnamefont
				{Yasunobu}\ \bibnamefont {Nakamura}}, \bibinfo {author} {\bibfnamefont
				{Achim}\ \bibnamefont {Marx}}, \bibinfo {author} {\bibfnamefont {Frank}\
				\bibnamefont {Deppe}}, \ and\ \bibinfo {author} {\bibfnamefont {Rudolf}\
				\bibnamefont {Gross}},\ }\bibfield  {title} {\enquote {\bibinfo {title}
				{Hysteretic flux response and nondegenerate gain of flux-driven {J}osephson
					parametric amplifiers},}\ }\href {\doibase 10.1103/PhysRevApplied.8.024012}
		{\bibfield  {journal} {\bibinfo  {journal} {Phys. Rev. Appl.}\ }\textbf
			{\bibinfo {volume} {8}},\ \bibinfo {pages} {024012} (\bibinfo {year}
			{2017})}\BibitemShut {NoStop}%
		\bibitem [{\citenamefont {Irastorza}\ and\ \citenamefont
			{Redondo}(2018)}]{irastorza2018new}%
		\BibitemOpen
		\bibfield  {author} {\bibinfo {author} {\bibfnamefont {Igor~G.}\ \bibnamefont
				{Irastorza}}\ and\ \bibinfo {author} {\bibfnamefont {Javier}\ \bibnamefont
				{Redondo}},\ }\bibfield  {title} {\enquote {\bibinfo {title} {New
					experimental approaches in the search for axion-like particles},}\ }\href
		{\doibase https://doi.org/10.1016/j.ppnp.2018.05.003} {\bibfield  {journal}
			{\bibinfo  {journal} {Prog. Part. Nucl. Phys.}\ }\textbf {\bibinfo {volume}
				{102}},\ \bibinfo {pages} {89--159} (\bibinfo {year} {2018})}\BibitemShut
		{NoStop}%
		\bibitem [{\citenamefont {Chaudhuri}\ \emph {et~al.}(2018)\citenamefont
			{Chaudhuri}, \citenamefont {Irwin}, \citenamefont {Graham},\ and\
			\citenamefont {Mardon}}]{chaudhuri2018fundamental}%
		\BibitemOpen
		\bibfield  {author} {\bibinfo {author} {\bibfnamefont {S.}~\bibnamefont
				{Chaudhuri}}, \bibinfo {author} {\bibfnamefont {K.}~\bibnamefont {Irwin}},
			\bibinfo {author} {\bibfnamefont {P.~W.}\ \bibnamefont {Graham}}, \ and\
			\bibinfo {author} {\bibfnamefont {J.}~\bibnamefont {Mardon}},\ }\bibfield
		{title} {\enquote {\bibinfo {title} {Fundamental limits of electromagnetic
					axion and hidden-photon dark matter searches: part {I} -- the quantum
					limit},}\ }\href {http://arxiv.org/abs/1803.01627} {\bibfield  {journal}
			{\bibinfo  {journal} {arXiv:1803.01627}\ } (\bibinfo {year}
			{2018})}\BibitemShut {NoStop}%
		\bibitem [{\citenamefont {Kenany}\ \emph {et~al.}(2017)\citenamefont {Kenany},
			\citenamefont {Anil}, \citenamefont {Backes}, \citenamefont {Brubaker},
			\citenamefont {Cahn}, \citenamefont {Carosi}, \citenamefont {Gurevich},
			\citenamefont {Kindel}, \citenamefont {Lamoreaux}, \citenamefont {Lehnert},
			\citenamefont {Lewis}, \citenamefont {Malnou}, \citenamefont {Palken},
			\citenamefont {Rapidis}, \citenamefont {Root}, \citenamefont {Simanovskaia},
			\citenamefont {Shokair}, \citenamefont {Urdinaran}, \citenamefont {van
				Bibber},\ and\ \citenamefont {Zhong}}]{alkenany2017design}%
		\BibitemOpen
		\bibfield  {author} {\bibinfo {author} {\bibfnamefont {S.~Al}\ \bibnamefont
				{Kenany}}, \bibinfo {author} {\bibfnamefont {M.A.}\ \bibnamefont {Anil}},
			\bibinfo {author} {\bibfnamefont {K.M.}\ \bibnamefont {Backes}}, \bibinfo
			{author} {\bibfnamefont {B.M.}\ \bibnamefont {Brubaker}}, \bibinfo {author}
			{\bibfnamefont {S.B.}\ \bibnamefont {Cahn}}, \bibinfo {author} {\bibfnamefont
				{G.}~\bibnamefont {Carosi}}, \bibinfo {author} {\bibfnamefont {Y.V.}\
				\bibnamefont {Gurevich}}, \bibinfo {author} {\bibfnamefont {W.F.}\
				\bibnamefont {Kindel}}, \bibinfo {author} {\bibfnamefont {S.K.}\ \bibnamefont
				{Lamoreaux}}, \bibinfo {author} {\bibfnamefont {K.W.}\ \bibnamefont
				{Lehnert}}, \bibinfo {author} {\bibfnamefont {S.M.}\ \bibnamefont {Lewis}},
			\bibinfo {author} {\bibfnamefont {M.}~\bibnamefont {Malnou}}, \bibinfo
			{author} {\bibfnamefont {D.A.}\ \bibnamefont {Palken}}, \bibinfo {author}
			{\bibfnamefont {N.M.}\ \bibnamefont {Rapidis}}, \bibinfo {author}
			{\bibfnamefont {J.R.}\ \bibnamefont {Root}}, \bibinfo {author} {\bibfnamefont
				{M.}~\bibnamefont {Simanovskaia}}, \bibinfo {author} {\bibfnamefont {T.M.}\
				\bibnamefont {Shokair}}, \bibinfo {author} {\bibfnamefont {I.}~\bibnamefont
				{Urdinaran}}, \bibinfo {author} {\bibfnamefont {K.A.}\ \bibnamefont {van
					Bibber}}, \ and\ \bibinfo {author} {\bibfnamefont {L.}~\bibnamefont
				{Zhong}},\ }\bibfield  {title} {\enquote {\bibinfo {title} {Design and
					operational experience of a microwave cavity axion detector for the 20-100
					$\mu$ev range},}\ }\href {\doibase
			https://doi.org/10.1016/j.nima.2017.02.012} {\bibfield  {journal} {\bibinfo
				{journal} {Nucl. Instr. Meth. Phys. Res. A}\ }\textbf {\bibinfo {volume}
				{854}},\ \bibinfo {pages} {11--24} (\bibinfo {year} {2017})}\BibitemShut
		{NoStop}%
		\bibitem [{\citenamefont {Brubaker}\ \emph
			{et~al.}(2017{\natexlab{b}})\citenamefont {Brubaker}, \citenamefont {Zhong},
			\citenamefont {Lamoreaux}, \citenamefont {Lehnert},\ and\ \citenamefont {van
				Bibber}}]{brubaker2017haystac}%
		\BibitemOpen
		\bibfield  {author} {\bibinfo {author} {\bibfnamefont {B.~M.}\ \bibnamefont
				{Brubaker}}, \bibinfo {author} {\bibfnamefont {L.}~\bibnamefont {Zhong}},
			\bibinfo {author} {\bibfnamefont {S.~K.}\ \bibnamefont {Lamoreaux}}, \bibinfo
			{author} {\bibfnamefont {K.~W.}\ \bibnamefont {Lehnert}}, \ and\ \bibinfo
			{author} {\bibfnamefont {K.~A.}\ \bibnamefont {van Bibber}},\ }\bibfield
		{title} {\enquote {\bibinfo {title} {{HAYSTAC} axion search analysis
					procedure},}\ }\href {\doibase 10.1103/PhysRevD.96.123008} {\bibfield
			{journal} {\bibinfo  {journal} {Phys. Rev. D}\ }\textbf {\bibinfo {volume}
				{96}},\ \bibinfo {pages} {123008} (\bibinfo {year}
			{2017}{\natexlab{b}})}\BibitemShut {NoStop}%
		\bibitem [{\citenamefont {Chung}(2016)}]{chung2016cultask}%
		\BibitemOpen
		\bibfield  {author} {\bibinfo {author} {\bibfnamefont {Woohyun}\ \bibnamefont
				{Chung}},\ }\bibfield  {title} {\enquote {\bibinfo {title} {{CULTASK, The
						Coldest Axion Experiment at CAPP/IBS in Korea}},}\ }\href
		{http://inspirehep.net/record/1497999?ln=en} {\bibfield  {journal} {\bibinfo
				{journal} {PoS}\ }\textbf {\bibinfo {volume} {CORFU2015}},\ \bibinfo {pages}
			{047} (\bibinfo {year} {2016})}\BibitemShut {NoStop}%
		\bibitem [{\citenamefont {McAllister}\ \emph {et~al.}(2017)\citenamefont
			{McAllister}, \citenamefont {Flower}, \citenamefont {Ivanov}, \citenamefont
			{Goryachev}, \citenamefont {Bourhill},\ and\ \citenamefont
			{Tobar}}]{mcallister2017organ}%
		\BibitemOpen
		\bibfield  {author} {\bibinfo {author} {\bibfnamefont {Ben~T.}\ \bibnamefont
				{McAllister}}, \bibinfo {author} {\bibfnamefont {Graeme}\ \bibnamefont
				{Flower}}, \bibinfo {author} {\bibfnamefont {Eugene~N.}\ \bibnamefont
				{Ivanov}}, \bibinfo {author} {\bibfnamefont {Maxim}\ \bibnamefont
				{Goryachev}}, \bibinfo {author} {\bibfnamefont {Jeremy}\ \bibnamefont
				{Bourhill}}, \ and\ \bibinfo {author} {\bibfnamefont {Michael~E.}\
				\bibnamefont {Tobar}},\ }\bibfield  {title} {\enquote {\bibinfo {title} {The
					{ORGAN} experiment: an axion haloscope above 15 {GH}z},}\ }\href
		{https://www.sciencedirect.com/science/article/pii/S2212686417300602}
		{\bibfield  {journal} {\bibinfo  {journal} {Phys. Dark Univ.}\ }\textbf
			{\bibinfo {volume} {18}},\ \bibinfo {pages} {67--72} (\bibinfo {year}
			{2017})}\BibitemShut {NoStop}%
		\bibitem [{\citenamefont {Ouellet}\ \emph {et~al.}(2019)\citenamefont
			{Ouellet}, \citenamefont {Salemi}, \citenamefont {Foster}, \citenamefont
			{Henning}, \citenamefont {Bogorad}, \citenamefont {Conrad}, \citenamefont
			{Formaggio}, \citenamefont {Kahn}, \citenamefont {Minervini}, \citenamefont
			{Radovinsky}, \citenamefont {Rodd}, \citenamefont {Safdi}, \citenamefont
			{Thaler}, \citenamefont {Winklehner},\ and\ \citenamefont
			{Winslow}}]{ouellet2018first}%
		\BibitemOpen
		\bibfield  {author} {\bibinfo {author} {\bibfnamefont {Jonathan~L.}\
				\bibnamefont {Ouellet}}, \bibinfo {author} {\bibfnamefont {Chiara~P.}\
				\bibnamefont {Salemi}}, \bibinfo {author} {\bibfnamefont {Joshua~W.}\
				\bibnamefont {Foster}}, \bibinfo {author} {\bibfnamefont {Reyco}\
				\bibnamefont {Henning}}, \bibinfo {author} {\bibfnamefont {Zachary}\
				\bibnamefont {Bogorad}}, \bibinfo {author} {\bibfnamefont {Janet~M.}\
				\bibnamefont {Conrad}}, \bibinfo {author} {\bibfnamefont {Joseph~A.}\
				\bibnamefont {Formaggio}}, \bibinfo {author} {\bibfnamefont {Yonatan}\
				\bibnamefont {Kahn}}, \bibinfo {author} {\bibfnamefont {Joe}\ \bibnamefont
				{Minervini}}, \bibinfo {author} {\bibfnamefont {Alexey}\ \bibnamefont
				{Radovinsky}}, \bibinfo {author} {\bibfnamefont {Nicholas~L.}\ \bibnamefont
				{Rodd}}, \bibinfo {author} {\bibfnamefont {Benjamin~R.}\ \bibnamefont
				{Safdi}}, \bibinfo {author} {\bibfnamefont {Jesse}\ \bibnamefont {Thaler}},
			\bibinfo {author} {\bibfnamefont {Daniel}\ \bibnamefont {Winklehner}}, \ and\
			\bibinfo {author} {\bibfnamefont {Lindley}\ \bibnamefont {Winslow}},\
		}\bibfield  {title} {\enquote {\bibinfo {title} {First results from
					{ABRACADABRA}-10 cm: {A} search for {Sub-$\ensuremath{\mu}\mathrm{eV}$} axion
					dark matter},}\ }\href {\doibase 10.1103/PhysRevLett.122.121802} {\bibfield
			{journal} {\bibinfo  {journal} {Phys. Rev. Lett.}\ }\textbf {\bibinfo
				{volume} {122}},\ \bibinfo {pages} {121802} (\bibinfo {year}
			{2019})}\BibitemShut {NoStop}%
		\bibitem [{\citenamefont {Silva-Feaver}\ \emph {et~al.}(2017)\citenamefont
			{Silva-Feaver}, \citenamefont {Chaudhuri}, \citenamefont {Cho}, \citenamefont
			{Dawson}, \citenamefont {Graham}, \citenamefont {Irwin}, \citenamefont
			{Kuenstner}, \citenamefont {Li}, \citenamefont {Mardon}, \citenamefont
			{Moseley}, \citenamefont {Mule}, \citenamefont {Phipps}, \citenamefont
			{Rajendran}, \citenamefont {Steffen},\ and\ \citenamefont
			{Young}}]{silvafeaver2017design}%
		\BibitemOpen
		\bibfield  {author} {\bibinfo {author} {\bibfnamefont {M.}~\bibnamefont
				{Silva-Feaver}}, \bibinfo {author} {\bibfnamefont {S.}~\bibnamefont
				{Chaudhuri}}, \bibinfo {author} {\bibfnamefont {H.}~\bibnamefont {Cho}},
			\bibinfo {author} {\bibfnamefont {C.}~\bibnamefont {Dawson}}, \bibinfo
			{author} {\bibfnamefont {P.}~\bibnamefont {Graham}}, \bibinfo {author}
			{\bibfnamefont {K.}~\bibnamefont {Irwin}}, \bibinfo {author} {\bibfnamefont
				{S.}~\bibnamefont {Kuenstner}}, \bibinfo {author} {\bibfnamefont
				{D.}~\bibnamefont {Li}}, \bibinfo {author} {\bibfnamefont {J.}~\bibnamefont
				{Mardon}}, \bibinfo {author} {\bibfnamefont {H.}~\bibnamefont {Moseley}},
			\bibinfo {author} {\bibfnamefont {R.}~\bibnamefont {Mule}}, \bibinfo {author}
			{\bibfnamefont {A.}~\bibnamefont {Phipps}}, \bibinfo {author} {\bibfnamefont
				{S.}~\bibnamefont {Rajendran}}, \bibinfo {author} {\bibfnamefont
				{Z.}~\bibnamefont {Steffen}}, \ and\ \bibinfo {author} {\bibfnamefont
				{B.}~\bibnamefont {Young}},\ }\bibfield  {title} {\enquote {\bibinfo {title}
				{Design overview of {DM} {R}adio pathfinder experiment},}\ }\href
		{https://ieeexplore.ieee.org/document/7750582} {\bibfield  {journal}
			{\bibinfo  {journal} {IEEE Trans. Appl. Supercond.}\ }\textbf {\bibinfo
				{volume} {27}},\ \bibinfo {pages} {1--4} (\bibinfo {year}
			{2017})}\BibitemShut {NoStop}%
		\bibitem [{\citenamefont {Majorovits}\ \emph {et~al.}(2017)\citenamefont
			{Majorovits} \emph {et~al.}}]{majorovits2017madmax}%
		\BibitemOpen
		\bibfield  {author} {\bibinfo {author} {\bibfnamefont {B.}~\bibnamefont
				{Majorovits}} \emph {et~al.} (\bibinfo {collaboration} {MADMAX interest
				group}),\ }\bibfield  {title} {\enquote {\bibinfo {title} {{MADMAX: A new
						road to axion dark matter detection}},}\ }\href
		{http://arxiv.org/abs/1712.01062} {\bibfield  {journal} {\bibinfo  {journal}
				{arXiv:1712.01062}\ } (\bibinfo {year} {2017})}\BibitemShut {NoStop}%
		\bibitem [{\citenamefont {Sliwa}\ \emph {et~al.}(2015)\citenamefont {Sliwa},
			\citenamefont {Hatridge}, \citenamefont {Narla}, \citenamefont {Shankar},
			\citenamefont {Frunzio}, \citenamefont {Schoelkopf},\ and\ \citenamefont
			{Devoret}}]{sliwa2015reconfigurable}%
		\BibitemOpen
		\bibfield  {author} {\bibinfo {author} {\bibfnamefont {K.~M.}\ \bibnamefont
				{Sliwa}}, \bibinfo {author} {\bibfnamefont {M.}~\bibnamefont {Hatridge}},
			\bibinfo {author} {\bibfnamefont {A.}~\bibnamefont {Narla}}, \bibinfo
			{author} {\bibfnamefont {S.}~\bibnamefont {Shankar}}, \bibinfo {author}
			{\bibfnamefont {L.}~\bibnamefont {Frunzio}}, \bibinfo {author} {\bibfnamefont
				{R.~J.}\ \bibnamefont {Schoelkopf}}, \ and\ \bibinfo {author} {\bibfnamefont
				{M.~H.}\ \bibnamefont {Devoret}},\ }\bibfield  {title} {\enquote {\bibinfo
				{title} {Reconfigurable {J}osephson circulator/directional amplifier},}\
		}\href {\doibase 10.1103/PhysRevX.5.041020} {\bibfield  {journal} {\bibinfo
				{journal} {Phys. Rev. X}\ }\textbf {\bibinfo {volume} {5}},\ \bibinfo {pages}
			{041020} (\bibinfo {year} {2015})}\BibitemShut {NoStop}%
		\bibitem [{\citenamefont {Macklin}\ \emph {et~al.}(2015)\citenamefont
			{Macklin}, \citenamefont {O'Brien}, \citenamefont {Hover}, \citenamefont
			{Schwartz}, \citenamefont {Bolkhovsky}, \citenamefont {Zhang}, \citenamefont
			{Oliver},\ and\ \citenamefont {Siddiqi}}]{macklin2015a}%
		\BibitemOpen
		\bibfield  {author} {\bibinfo {author} {\bibfnamefont {C.}~\bibnamefont
				{Macklin}}, \bibinfo {author} {\bibfnamefont {K.}~\bibnamefont {O'Brien}},
			\bibinfo {author} {\bibfnamefont {D.}~\bibnamefont {Hover}}, \bibinfo
			{author} {\bibfnamefont {M.~E.}\ \bibnamefont {Schwartz}}, \bibinfo {author}
			{\bibfnamefont {V.}~\bibnamefont {Bolkhovsky}}, \bibinfo {author}
			{\bibfnamefont {X.}~\bibnamefont {Zhang}}, \bibinfo {author} {\bibfnamefont
				{W.~D.}\ \bibnamefont {Oliver}}, \ and\ \bibinfo {author} {\bibfnamefont
				{I.}~\bibnamefont {Siddiqi}},\ }\bibfield  {title} {\enquote {\bibinfo
				{title} {A near--quantum-limited {J}osephson traveling-wave parametric
					amplifier},}\ }\href {\doibase 10.1126/science.aaa8525} {\bibfield  {journal}
			{\bibinfo  {journal} {Science}\ }\textbf {\bibinfo {volume} {350}},\ \bibinfo
			{pages} {307--310} (\bibinfo {year} {2015})}\BibitemShut {NoStop}%
		\bibitem [{\citenamefont {Chapman}\ \emph {et~al.}(2017)\citenamefont
			{Chapman}, \citenamefont {Rosenthal}, \citenamefont {Kerckhoff},
			\citenamefont {Moores}, \citenamefont {Vale}, \citenamefont {Mates},
			\citenamefont {Hilton}, \citenamefont {Lalumi\`ere}, \citenamefont {Blais},\
			and\ \citenamefont {Lehnert}}]{chapman2017widely}%
		\BibitemOpen
		\bibfield  {author} {\bibinfo {author} {\bibfnamefont {Benjamin~J.}\
				\bibnamefont {Chapman}}, \bibinfo {author} {\bibfnamefont {Eric~I.}\
				\bibnamefont {Rosenthal}}, \bibinfo {author} {\bibfnamefont {Joseph}\
				\bibnamefont {Kerckhoff}}, \bibinfo {author} {\bibfnamefont {Bradley~A.}\
				\bibnamefont {Moores}}, \bibinfo {author} {\bibfnamefont {Leila~R.}\
				\bibnamefont {Vale}}, \bibinfo {author} {\bibfnamefont {J.~A.~B.}\
				\bibnamefont {Mates}}, \bibinfo {author} {\bibfnamefont {Gene~C.}\
				\bibnamefont {Hilton}}, \bibinfo {author} {\bibfnamefont {Kevin}\
				\bibnamefont {Lalumi\`ere}}, \bibinfo {author} {\bibfnamefont {Alexandre}\
				\bibnamefont {Blais}}, \ and\ \bibinfo {author} {\bibfnamefont {K.~W.}\
				\bibnamefont {Lehnert}},\ }\bibfield  {title} {\enquote {\bibinfo {title}
				{Widely tunable on-chip microwave circulator for superconducting quantum
					circuits},}\ }\href {\doibase 10.1103/PhysRevX.7.041043} {\bibfield
			{journal} {\bibinfo  {journal} {Phys. Rev. X}\ }\textbf {\bibinfo {volume}
				{7}},\ \bibinfo {pages} {041043} (\bibinfo {year} {2017})}\BibitemShut
		{NoStop}%
		\bibitem [{\citenamefont {Clerk}\ \emph {et~al.}(2010)\citenamefont {Clerk},
			\citenamefont {Devoret}, \citenamefont {Girvin}, \citenamefont {Marquardt},\
			and\ \citenamefont {Schoelkopf}}]{clerk2010introduction}%
		\BibitemOpen
		\bibfield  {author} {\bibinfo {author} {\bibfnamefont {A.~A.}\ \bibnamefont
				{Clerk}}, \bibinfo {author} {\bibfnamefont {M.~H.}\ \bibnamefont {Devoret}},
			\bibinfo {author} {\bibfnamefont {S.~M.}\ \bibnamefont {Girvin}}, \bibinfo
			{author} {\bibfnamefont {Florian}\ \bibnamefont {Marquardt}}, \ and\ \bibinfo
			{author} {\bibfnamefont {R.~J.}\ \bibnamefont {Schoelkopf}},\ }\bibfield
		{title} {\enquote {\bibinfo {title} {Introduction to quantum noise,
					measurement, and amplification},}\ }\href {\doibase
			10.1103/RevModPhys.82.1155} {\bibfield  {journal} {\bibinfo  {journal} {Rev.
					Mod. Phys.}\ }\textbf {\bibinfo {volume} {82}},\ \bibinfo {pages}
			{1155--1208} (\bibinfo {year} {2010})}\BibitemShut {NoStop}%
		\bibitem [{\citenamefont {Dicke}(1946)}]{dicke1946the}%
		\BibitemOpen
		\bibfield  {author} {\bibinfo {author} {\bibfnamefont {R.~H.}\ \bibnamefont
				{Dicke}},\ }\bibfield  {title} {\enquote {\bibinfo {title} {The measurement
					of thermal radiation at microwave frequencies},}\ }\href {\doibase
			10.1063/1.1770483} {\bibfield  {journal} {\bibinfo  {journal} {Rev. Sci.
					Instrum.}\ }\textbf {\bibinfo {volume} {17}},\ \bibinfo {pages} {268--275}
			(\bibinfo {year} {1946})}\BibitemShut {NoStop}%
		\bibitem [{\citenamefont {Kim}(1979)}]{kim1979weak}%
		\BibitemOpen
		\bibfield  {author} {\bibinfo {author} {\bibfnamefont {Jihn~E.}\ \bibnamefont
				{Kim}},\ }\bibfield  {title} {\enquote {\bibinfo {title} {Weak-interaction
					singlet and strong $\mathrm{CP}$ invariance},}\ }\href {\doibase
			10.1103/PhysRevLett.43.103} {\bibfield  {journal} {\bibinfo  {journal} {Phys.
					Rev. Lett.}\ }\textbf {\bibinfo {volume} {43}},\ \bibinfo {pages} {103--107}
			(\bibinfo {year} {1979})}\BibitemShut {NoStop}%
		\bibitem [{\citenamefont {Shifman}\ \emph {et~al.}(1980)\citenamefont
			{Shifman}, \citenamefont {Vainshtein},\ and\ \citenamefont
			{Zakharov}}]{shifman1980can}%
		\BibitemOpen
		\bibfield  {author} {\bibinfo {author} {\bibfnamefont {M.A.}\ \bibnamefont
				{Shifman}}, \bibinfo {author} {\bibfnamefont {A.I.}\ \bibnamefont
				{Vainshtein}}, \ and\ \bibinfo {author} {\bibfnamefont {V.I.}\ \bibnamefont
				{Zakharov}},\ }\bibfield  {title} {\enquote {\bibinfo {title} {Can
					confinement ensure natural {CP} invariance of strong interactions?}}\ }\href
		{\doibase https://doi.org/10.1016/0550-3213(80)90209-6} {\bibfield  {journal}
			{\bibinfo  {journal} {Nucl. Phys. B}\ }\textbf {\bibinfo {volume} {166}},\
			\bibinfo {pages} {493--506} (\bibinfo {year} {1980})}\BibitemShut {NoStop}%
		\bibitem [{\citenamefont {Kurpiers}\ \emph {et~al.}(2017)\citenamefont
			{Kurpiers}, \citenamefont {Walter}, \citenamefont {Magnard}, \citenamefont
			{Salathe},\ and\ \citenamefont {Wallraff}}]{kurpiers2017characterizing}%
		\BibitemOpen
		\bibfield  {author} {\bibinfo {author} {\bibfnamefont {Philipp}\ \bibnamefont
				{Kurpiers}}, \bibinfo {author} {\bibfnamefont {Theodore}\ \bibnamefont
				{Walter}}, \bibinfo {author} {\bibfnamefont {Paul}\ \bibnamefont {Magnard}},
			\bibinfo {author} {\bibfnamefont {Yves}\ \bibnamefont {Salathe}}, \ and\
			\bibinfo {author} {\bibfnamefont {Andreas}\ \bibnamefont {Wallraff}},\
		}\bibfield  {title} {\enquote {\bibinfo {title} {Characterizing the
					attenuation of coaxial and rectangular microwave-frequency waveguides at
					cryogenic temperatures},}\ }\href {\doibase 10.1140/epjqt/s40507-017-0059-7}
		{\bibfield  {journal} {\bibinfo  {journal} {EPJ Quantum Technol.}\ }\textbf
			{\bibinfo {volume} {4}},\ \bibinfo {pages} {8} (\bibinfo {year}
			{2017})}\BibitemShut {NoStop}%
		\bibitem [{Note1()}]{Note1}%
		\BibitemOpen
		\bibinfo {note} {In a real haloscope search, since the local oscillator being
			used for homodyne measurement would be stepped along with the cavity,
			non-axion-induced power excesses in the rf are trickier to reject than their
			IF counterparts \cite {brubaker2017haystac}.}\BibitemShut {Stop}%
		\bibitem [{\citenamefont {Savitzky}\ and\ \citenamefont
			{Golay}(1964)}]{savitzky1964smoothing}%
		\BibitemOpen
		\bibfield  {author} {\bibinfo {author} {\bibfnamefont {Abraham.}\
				\bibnamefont {Savitzky}}\ and\ \bibinfo {author} {\bibfnamefont {M.~J.~E.}\
				\bibnamefont {Golay}},\ }\bibfield  {title} {\enquote {\bibinfo {title}
				{Smoothing and differentiation of data by simplified least squares
					procedures.}}\ }\href {\doibase 10.1021/ac60214a047} {\bibfield  {journal}
			{\bibinfo  {journal} {Anal. Chem.}\ }\textbf {\bibinfo {volume} {36}},\
			\bibinfo {pages} {1627--1639} (\bibinfo {year} {1964})}\BibitemShut {NoStop}%
	\end{thebibliography}
	
	%

\end{document}